\documentclass[a4paper,fleqn,usenatbib]{mnras}

\usepackage{newtxtext,newtxmath}
\usepackage{multirow}

\usepackage[T1]{fontenc}
\usepackage{ae,aecompl}

\usepackage{graphicx}	
\usepackage{amsmath}	
\usepackage{amssymb}	

\newcommand{\elecd}{$n_{\rm e}$} 
\newcommand{\elect}{$T_{\rm e}$}

\newcommand{\foii}{[O\thinspace{\sc ii}]} 
\newcommand{\foiii}{[O\thinspace{\sc iii}]} 
\newcommand{\fsii}{[S\thinspace{\sc ii}]} 
\newcommand{\fsiii}{[S\thinspace{\sc iii}]} 
 
\newcommand{\fnii}{[N\thinspace{\sc ii}]}

\newcommand{\oiii}{O\thinspace{\sc iii}}

\newcommand{\oii}{O\thinspace{\sc ii}}

\newcommand{\hi}{H\,{\sc i}} 
\newcommand{\hii}{H\thinspace{\sc ii}} 
 
\newcommand{\hei}{He\thinspace{\sc i}} 
\newcommand{\heii}{He\thinspace{\sc ii}}

\newcommand\ionic[2]{${\rm #1^{#2}}$}  

\newcommand{\cmc}{{\rm cm$^{-3}$}} 

\newcommand{\lamb}{$\lambda$}

\title[Helium abundances and gradient in Galactic nebulae]{Helium abundances and its radial gradient from the spectra of {\hii} regions and ring nebulae of the Milky Way}

\author[J. E. M\'endez-Delgado et al.]
       {J. E. M\'endez-Delgado $^{1,2}$\thanks{E-mail: jemd@iac.es}, 
        C. Esteban$^{1,2}$, J. Garc{\'{\i}}a-Rojas$^{1,2}$, K. Z. Arellano-C\'ordova$^{1,3}$ 
        \newauthor 
        and M. Valerdi$^{4}$\\
\\
	$^1$Instituto de Astrof\'\i sica de Canarias, E-38200 La Laguna, Tenerife, Spain\\
        $^2$Departamento de Astrof\'\i sica, Universidad de La Laguna, E-38206, La Laguna, Tenerife, Spain\\
        $^3$Instituto Nacional de Astrof\'\i sica, \'Optica y Electr\'onica. Apdo. Postal 51 y 216, Puebla, Mexico\\
        $^4$Instituto de Astronom\'ia, Universidad Nacional Aut\'onoma de M\'exico, Apdo. Postal 70-264 Ciudad Universitaria, Mexico}

\date{Accepted XXX. Received YYY; in original form ZZZ}

\pubyear{2015}

\begin{document}
\label{firstpage}
\pagerange{\pageref{firstpage}--\pageref{lastpage}}
\maketitle

\begin{abstract}
We determine the radial abundance gradient of helium in the disc of the Galaxy from published spectra of 19 {\hii} regions and ring nebulae surrounding massive O stars. We revise the Galactocentric distances of the objects considering {\it Gaia} DR2 parallaxes and determine the physical conditions and the ionic abundance of \ionic{He}{+} in a homogeneous way, using between 3 and 10 {\hei} recombination lines in each object. We estimate the total He abundance of the nebulae and its radial abundance gradient using four different ICF(He) schemes. The slope of the gradient is always negative and weakly dependent on the ICF(He) scheme, especially when only the objects with log($\eta$) $<$ 0.9 are considered. The slope values go from $-$0.0078 to $-$0.0044 dex kpc$^{-1}$, consistent with the predictions of chemical evolution models of  the Milky Way and chemodynamical simulations of disc galaxies. Finally, we estimate the abundance deviations of He, O and N in a sample of ring nebulae around Galactic WR stars, finding a quite similar He overabundance  of about +0.24 $\pm$ 0.11 dex in three stellar ejecta ring nebulae.
\end{abstract}




\section{Introduction} \label{intro}

Although rare on Earth, helium is the second most abundant element in the Universe and constitutes about 24-25\% of its baryonic mass. The vast majority of the cosmic helium was produced during the primordial nucleosynthesis phase just after the Big Bang. The fraction of primordial mass in helium, $Y_P$, has been determined following three different techniques. The results of {\it Wilkinson Microwave Anisotropy Probe} (WMAP) and Planck satellites devoted to the study of the Cosmic Microwave Background (CMB) anisotropies have obtained values between 0.245 and 0.247 assuming the Standard Big Bang Nucleosynthesis \citep{cocetal05, planck18}. Studies of intergalactic clouds near pristine absorption systems along the line of sight of bright distant quasars determine upper limits of $Y_P$ in the range 0.225 and 0.283 \citep{cookefumagali18}. The last technique, based on the analysis of spectra of {\hii} regions of metal-poor galaxies gives values of $Y_P$ between 0.238 and 0.257 in the most recent works \citep{izotovetal14,averetal15,apeimbertetal16,fernandezetal18,valerdietal19}. 

After the Big Bang, helium is produced by hydrostatic nucleosynthesis in the interior of stars of all initial masses. Low-mass stars produce this element through the proton-proton chain while intermediate mass and massive ones via the CNO cycle. Helium can be also efficiently destroyed in stellar interiors by the triple-alpha process. The amount of this element that is actually ejected by a given star and enrich the ISM depends on its initial mass and the importance of stellar winds.   

The analysis of Galactic {\hii} region spectra indicates the presence of radial gradients of the abundances of heavy elements -- such as O, N, Ne, S, Ar or Cl -- along the disc of the Milky Way \citep[e.g.][]{shaveretal83, deharvengetal00, rudolphetal06, balseretal11, estebangarciarojas18}.  The form of such gradients reflects the action of stellar nucleosynthesis, the distribution and history of star formation and gas flows 
in the chemical evolution of the Galaxy. Although the helium abundance should increase with the metallicity, there is not a clear evidence of the presence of a radial gradient of this element in the Milky Way. Some authors \citep[e.g.][]{peimbertetal78,talentdufour79} find a slight or marginal evidence of a negative gradient but others find a flat distribution of the helium abundance along the Galactic disc \citep[e.g.][see \S\ref{sec:gradHe} for further discussion]{shaveretal83, fernandezmartinetal17}. 

There are several sources of uncertainty in the determination of the total abundance of helium. The most important one is related to the ionization structure of the nebulae. The spectrum of normal {\hii} regions only 
shows recombination lines of \ionic{He}{+}. Since the ionization potential of \ionic{He}{0} is 24.6 eV, we expect the presence of neutral helium in the {\hii} region, but it cannot be observed. To determine the total helium 
abundance from the measured \ionic{He}{+}/\ionic{H}{+} ratio, we have to rely on an ionization correction factor, ICF. In the absence of a tailored photoionization model  for the object, we have to assume a particular ICF scheme based on the results of grids of photoionization models or on the similarity of ionization potentials of other particular ions. All ICF schemes are parameterized by the ionization degree measured in the nebular spectrum. There are several ICF schemes for helium available in the literature and based on different ionic ratios \citep[e.g.][]{peimberttorrespeimbert77,peimbertetal92,kunthsargent83,zhangliu03}. 

There are two more additional sources of uncertainty in the determination of the total abundance of helium and they are related to deviations from the pure recombination spectrum of {\hei}.   
The {\hei} atom has two different level states depending on its total spin quantum number, singlets and triplets. While -- in principle -- the intensity ratios of singlet lines should follow the predictions of recombination theory, the triplet spectrum is affected by the metastability of the lowest triplet level, the 2$^3$S \citep{osterbrockferland06}. 
This metastable level produces important collisional and and self-absorption effects. Collisions with free electrons may pump electrons from 2$^3$S to upper levels, enhancing the intensities of the lines coming from those excited levels with respect to the predictions of recombination theory. Although triplet lines are the most affected, some singlets lines can be also enhanced \citep{saweyberrington93, kingdonferland95}. Photons of transitions ending on 2$^3$S can be reabsorbed, and, eventually, emitted as other {\hei} lines. This self-absorption process increases or decreases line intensities of triplet lines with respect to recombination according to the line considered. The strength of collisional effects on the {\hei} spectrum depends on the electron density and temperature of the ionized gas, being higher in denser and hotter nebulae. The calculations of the recombination spectrum of {\hei} performed by \citet{porteretal12} include the collisional effects in the level populations of \ionic{He}{0}. Self-absorption effects depend mostly on density, and are not expected to be very important in {\hii} regions \citep{benjaminetal02}. 

The aim of this paper is to explore the existence of a radial gradient of helium in the Milky Way using the best dataset of deep spectra of {\hii} regions available to date and considering revised distances for the nebulae. The selection gives priority to deep spectra that have been treated homogeneously in their reduction process. The structure of this paper is as follows. In \S\ref{sec:data} we describe the sample of {\hii} region spectra used and the determination of the revised distances. In \S\ref{results} we present the values of physical conditions, electron density and temperature -- {\elecd} and {\elect} -- adopted for each object; the  \ionic{He}{+} and He abundance recalculations, as well as discuss the different ICFs used. In \S\ref{sec:gradHe} we discuss the results pertaining to the Galactic He radial gradients. In \S\ref{sec:HeWR} we compute and discuss the abundance deviations of He, O and N with respect to the radial abundance gradients shown by a sample of Galactic ring nebulae around WR stars. Finally, in \S\ref{conclusions} we summarize our main conclusions.

\section{Observational data and revised distances} 
\label{sec:data}

We have used reddening corrected intensity ratios of several emission lines of 38 spectra corresponding to 24 Galactic {\hii} regions and ring nebulae around massive stars (Wolf-Rayet, WR or O-type stars). The data have been obtained, reduced and analyzed by our group and published in \citet{estebanetal04, estebanetal13, estebanetal16, estebanetal17}, \citet{estebangarciarojas18} and  \citet{garciarojasetal04, garciarojasetal05, garciarojasetal06, garciarojasetal07}. We will refer this group of publications as the "set of source papers''. The spectra were obtained with the Ultraviolet Visual Echelle Spectrograph 
\citep[UVES,][]{dodoricoetal00} at the Very Large Telescope (VLT); the OSIRIS (Optical System for Imaging and low-Intermediate-Resolution Integrated Spectroscopy) spectrograph \citep{cepaetal00, cepaetal03} at the 10.4 m Gran Telescopio Canarias (GTC) and the Magellan Echellette (MagE) spectrograph at the 6.5m Clay Telescope \citep{marshalletal08}. The details of the observations and the instrument configurations used are described in the set of source papers. 

In the set of source papers the assumed distance of the Sun to the Galactic Centre was $R_{\rm 0}$ = 8.0 kpc \citep{reid93}. Most of the recent determinations do not provide substantially different values of this parameter, but the precision has increased considerably. \citet{bland-hawthorngerhard} present a compilation of direct (primary), model-based and secondary determinations, proposing a best estimate for the distance to the Galactic Center of 8.2 $\pm$ 0.1 kpc. We adopt this value and its uncertainty for all our calculations of the Galactocentric distance ($R_{\rm G}$) of the objects.  The GRAVITY collaboration has published a very recent geometric distance determination of the Galactic centre black hole with 0.3\% uncertainty \citep{gravity19}, which is entirely consistent with 8.2 kpc.  

The set of source papers give the Galactocentric distances, $R_{\rm G}$ for each object.  In the case of the H~{\sc ii} regions, \citet{estebanetal17} and  \citet{estebangarciarojas18} adopted the mean values of kinematic and stellar distances given in different published references and an uncertainty corresponding to their standard deviation. In order to obtain an improved set of $R_{\rm G}$ values, we have made a revision taking into account new distance measurements based on {\it Gaia} parallaxes of the second data release (DR2). For M8, M16, M17, M20, M42 and NGC 3576 we have adopted the distances derived  by \citet{binderpovich18} from {\it Gaia} parallaxes, with typical uncertainties in the range 0.1-0.3 kpc. For the rest of the objects we have searched the distances of the ionizing and/or associated star (or stars) inferred by \citet{bailer-jonesetal18} from {\it Gaia} DR2 data and a distance prior that varies smoothly as a function of the Galactic coordinates of the objects according to a Galaxy model. In general, we find that distances based on {\it Gaia} parallaxes are very similar to those assumed previously in \citet{estebanetal17} and  \citet{estebangarciarojas18}. The mean difference for the 23 objects for which the two kinds of determinations can be compared is about 9\% (the median is 5\%). The uncertainties in the distance of the {\it Gaia} parallaxes tend to be larger than about 1 kpc and larger than those quoted by \citet{estebanetal17} and  \citet{estebangarciarojas18} for objects at heliocentric distances about or larger than 5 kpc. The largest differences  correspond to the the most external object of the sample (Sh~2-209, 38\%), but also for the objects located at $R_{\rm G}$ $\leq$ 7 kpc, with a mean difference of about 16\%. 

In the case of the ring nebulae (G2.4+1.4, NGC 6888, NGC 7535, RCW 52, RCW 58, Sh 2-298 and Sh 2-308), all of them are ionized by a single well identified WR or O star and are located at heliocentric distances smaller than 5 kpc. Their distances have been taken directly from the database  of \citet{bailer-jonesetal18}. For the rest of the objects -- for which we do not have a single ionized source or lack information about their ionizing sources -- we compare the distances obtained from {\it Gaia} DR2 data \citep{bailer-jonesetal18} with those adopted by \citet{estebanetal17} or \citet{estebangarciarojas18}, assuming one or the other depending on which determination gives the smallest uncertainties. We describe the details for each object below. 

{\it NGC~2579}. We analyzed the {\it Gaia} DR2 data for stars located within 1 arcmin around the center of the {\hii} region. The parallaxes of the stars give a large and rather  uncertain heliocentric distance for the object (5.31 $\pm$ 1.01 kpc), that corresponds to an average $R_{\rm G}$ of about 10.88 $\pm$ 0.78 kpc. \citet{copettietal07} carried out a detailed study of spectroscopic parallaxes of the possible ionizing stars of the nebula, finding a somewhat larger distance of $R_{\rm G}$  = 12.40 $\pm$ 0.50 kpc. This value is consistent with the kinematic distance derived by those authors from the H$\alpha$ velocity field as well as the previous photometrical and kinematical determinations by \citet{russeiletal07}. We assume the distance determined by \citet{copettietal07} for this object. 

{\it Sh 2-83}. There are not specific studies investigating the ionizing sources of this nebula. The only studies that give the distance of Sh~2-83 are those by \citet{caplanetal00} and \citet{andersonetal15} based on kinematic data of the nebular gas, and both give the same  heliocentric distance of 18.4 kpc, the largest value of the whole sample. If the object is at such large distance, its parallax would be too small to be well measured in {\it Gaia} DR2. Therefore, it is not surprising that we do not find stars at so large distances in our analysis for stars located within 1 arcmin around the center of the {\hii} region. We adopted the distance obtained by \citet{caplanetal00} and \citet{andersonetal15}, that was the one assumed by \citet{estebanetal17} for this object. 

 {\it Sh 2-209}. \citet{chiniwink84} identified 3 ionizing stars for this nebula, indicating their position in a photograph. We have searched for those stars in the {\it Gaia} DR2 database obtaining $R_{\rm G}$ about 10.5 kpc for all of them. These values are very much lower than the 17.00 $\pm$ 0.70 kpc adopted by \citet{estebanetal17} based on several different but consistent kinematical and spectrophotometrical determinations. This is a serious drawback. The rather low O/H ratio of Sh 2-209 is more consistent with the larger value of the distance considering the predictions of the latest determinations of the Galactic O/H gradient \citep[e.g.][see their Figure 11]{estebanetal17} for $R_{\rm G}$ $\sim$ 10.5 kpc. Conversely, If we accept that the object is at $R_{\rm G}$ = 17 kpc, it would imply a heliocentric distance of about 9 kpc, but we do not find stars located at such large distances within 3 arcmin around  the centre of the {\hii} region in {\it Gaia} DR2. Such a high heliocentric distance may imply a parallax too small to be measured in {\it Gaia} DR2 and this may be the explanation of not finding suitable distant stars in the catalogue. We finally adopt the distance of 17.00 $\pm$ 0.70 kpc given by \citet{estebanetal17} for this object. 

 {\it Sh 2-311}. This {\hii} region is ionized by a single star: HD 64315, but it is a multiple stellar system \citep{lorenzoetal17} and its {\it Gaia} parallax is negative. We obtained and represented the parallaxes and proper motions of all stars located within 5 arcmin around HD~64315 (1810 stars) finding 3 defined peaks in the distribution. We calculated the distance corresponding to the center of each peak and the uncertainty corresponding to 1$\sigma$. We assumed the distance of 11.22 $\pm$ 0.22 kpc corresponding to the peak of the most distant distribution of stars, which is very similar to that of 11.1 $\pm$ 0.4 kpc assumed by \citet{estebanetal17}. We assume the value determined from the {\it Gaia} data for this object
 
 The rest of the sample objects are located at heliocentric distances about or larger than 5 kpc and their {\it Gaia} DR2 parallaxes give rather uncertain distances. We have finally assumed the distance adopted by \citet{estebanetal17} or \citet{estebangarciarojas18} for all these nebulae. As it has been said before, this distance corresponds to the mean of several independent photometrical and kinematical determinations \citep[obtained mainly from][]{caplanetal00,russeil03,russeiletal07,quirezaetal06, balseretal11,fosterbrunt15} that provide a standard deviation lower than the distance uncertainty obtained from the {\it Gaia} parallaxes.

 {\it NGC~3603}.  This distant object has been cited as a Galactic giant {\hii} region and compared with 30 Doradus in the Large Magellanic Cloud. \citet{drewetal19} compiled a list of almost 300 candidate O stars of the associated cluster and estimate an heliocentric distance of 7.0 $\pm$ 1.0 kpc from the {\it Gaia} DR2 data, that implies a $R_{\rm G}$ = 8.6 $\pm$ 1.0 kpc, in complete agreement with the distance of 8.6 $\pm$ 0.4 kpc adopted by \citet{estebanetal17}. 
 
{\it Sh 2-100}. \citet{samaletal10} found that Sh~2-100 is in a molecular complex containing 7 {\hii} regions. Those authors identify several ionizing stars inside the complex. We have obtained their {\it Gaia} parallaxes that give a very large heliocentric distance of  9.5$^{+2.4}_{-1.7}$ kpc for the object, corresponding to $R_{\rm G}$ = 10.2$^{+2.4}_{-1.7}$ kpc, which is consistent with the value adopted by \citet{estebanetal17} but much more uncertain.

{\it Sh 2-127}. \citet{rudolphetal96} conclude that this {\hii} region is ionized by two O-type stars. Only one of them is visible in the optical, that seems to be associated to the {\it Gaia} DR2 source  217611164377605888. The parallax of this source is very uncertain and gives a very large heliocentric distance of 9.5$^{+2.6}_{-1.9}$ kpc, that corresponds to $R_{\rm G}$ = 13.2$^{+2.1}_{-1.6}$ kpc. This value is consistent with the distance of 14.2 $\pm$ 1.0 kpc quoted by \citet{estebanetal17} within the errors. 

{\it Sh 2-128}. This {\hii} region is ionized by the star ALS 19702 \citep{bohigastapia03}. Its parallax is very uncertain, due to its heliocentric distance of 8.6$^{+2.2}_{-1.6}$ kpc, corresponding to $R_{\rm G}$ = 12.7$^{+1.8}_{-1.3}$ kpc, but entirely consistent with that of 12.50 $\pm$ 0.40 kpc quoted by \citet{estebanetal17}. 

{\it Sh 2-152}. We use the star no. 4 of \citet{russeiletal07} for obtaining the {\it Gaia} DR2 parallax (the stars no. 1 to 3 are more likely associated with Sh~2-153), obtaining a $R_{\rm G}$ of 10.9$^{+1.6}_{-1.1}$ kpc, in agreement but slightly more uncertain than the distance of 10.3 $\pm$ 1.0 kpc given by \citet{estebangarciarojas18}. 

{\it Sh 2-212}. \citet{moffatetal79} catalogued several stars associated to this star-forming region. Their {\it Gaia} parallaxes are very uncertain and provide an heliocentric distance of 6.0$^{+2.2}_{-1.5}$ kpc and $R_{\rm G}$ = 13.9$^{+2.2}_{-1.5}$ kpc, which is consistent with that of 14.6 $\pm$ 1.4 kpc given by \citet{estebanetal17}.

 {\it Sh 2-288}. \citet{avedisovakondratenko84} found that this nebulae is ionized by the star GSC 04823-00146. Its very uncertain parallax gives $R_{\rm G}$ = 12.7 $^{+1.5}_{-1.1}$ kpc. We use the more precise distance of 14.10 $\pm$ 0.40 kpc quoted by \citet{estebanetal17}. 

\begin{figure} 
\centering \includegraphics[scale=0.70]{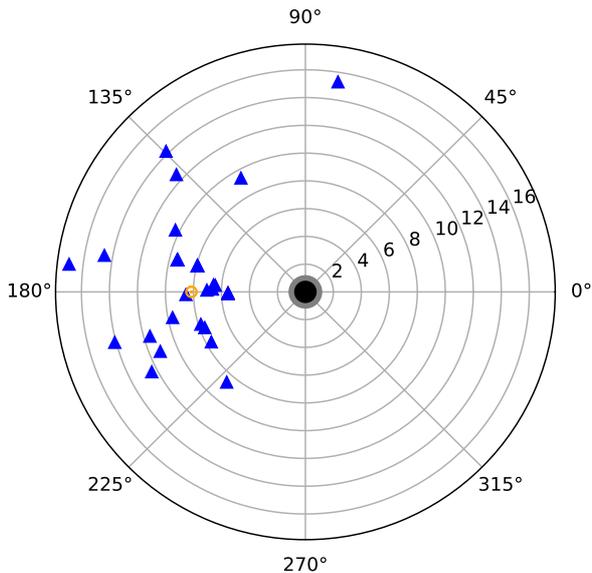} 
\caption{Spatial distribution of the sample nebulae (blue triangles) onto the Galactic plane with respect to the centre of the Milky Way. The concentric circles indicate increasing Galactocentric distances (in kpc). The Sun position is given by the orange small circle. } 
\label{fig:dist_map} 
\end{figure}

In Table~\ref{tab:sample} we list the objects  whose spectra are used in this paper along with their type ({\hii} region or ring nebula around WR or O-type star), adopted $R_{\rm G}$, instrument and telescope with which they were observed and the reference to their published spectra. In Fig.~\ref{fig:dist_map} we show the spatial distribution of the sample nebulae onto the Galactic plane. 

\begin{table*}
\centering
\caption{Sample of objects which spectroscopical data are used in this paper.} 
\label{tab:sample} 
\begin{tabular}{lccccc}
\hline
Object & Type$^{\rm a}$ & $R_{\rm G}$ (kpc)&Telescope & Instrument& Reference \\ \hline 
\multirow{2}{*}{G2.4+1.4} & \multirow{2}{*}{WR-RN} & \multirow{2}{*}{$5.53 ^{+0.32} _{-0.29}$$^{\rm b}$} &GTC & OSIRIS & \multirow{2}{*}{\citet{estebanetal16}}\\ 
 & & & MTC & MagE& \\ 
M8 & {\hii} & 7.04 $\pm$ 0.20$^{\rm b}$&VLT & UVES&  \citet{garciarojasetal07}\\ 
M16 & {\hii} & 6.58 $\pm$ 0.28$^{\rm b}$&VLT & UVES & \citet{garciarojasetal06}\\ 
M17 & {\hii} & 6.46 $\pm$ 0.26$^{\rm b}$&VLT & UVES& \citet{garciarojasetal07} \\ 
M20 & {\hii} & 6.64 $\pm$ 0.31$^{\rm b}$&VLT & UVES&  \citet{garciarojasetal06}\\ 
M42 & {\hii} & 8.54 $\pm$ 0.11$^{\rm b}$&VLT & UVES& \citet{estebanetal04}\\
NGC 2579 & {\hii} & 12.40$\pm$0.50$^{\rm c}$ &VLT & UVES &\citet{estebanetal13}\\ 
NGC 3576 & {\hii} & 7.64 $\pm$ 0.30$^{\rm b}$& VLT & UVES& \citet{garciarojasetal04}\\ 
NGC 3603 & {\hii} & 8.60 $\pm$ 0.40$^{\rm c}$ &VLT & UVES&  \citet{garciarojasetal06}\\ 
NGC 6888 &  WR-RN & 7.94 $\pm$ 0.16$^{\rm b}$ &GTC & OSIRIS &\citet{estebanetal16}\\ 
NGC 7635 & O-RN & 9.43$^{+0.21} _{-0.20}$$^{\rm b}$ &GTC & OSIRIS& \citet{estebanetal16} \\ 
RCW 52 & O-RN & 7.83$^{+0.21} _{-0.10}$$^{\rm b}$ &MTC & MagE &\citet{estebanetal16}\\ 
RCW 58 & WR-RN & 7.63$^{+0.50} _{-0.42}$$^{\rm b}$ &MTC & MagE& \citet{estebanetal16}\\ 
Sh~2-83 & {\hii} & 15.30$\pm$0.10$^{\rm c}$ &GTC & OSIRIS& \citet{estebanetal17}\\ 
Sh~2-100 & {\hii} & 9.40$\pm$0.30$^{\rm c}$ &GTC & OSIRIS&\citet{estebanetal17} \\ 
Sh~2-127 & {\hii} & 14.2$\pm$1.0$^{\rm c}$ &GTC & OSIRIS& \citet{estebanetal17}\\ 
Sh~2-128 & {\hii} & 12.50$\pm$0.40$^{\rm c}$ &GTC & OSIRIS&\citet{estebanetal17} \\ 
Sh~2-152 & {\hii} & 10.3$\pm$1.0$^{\rm c}$ &GTC & OSIRIS &\citet{estebangarciarojas18} \\  
Sh~2-209 & {\hii} & 17.00$\pm$0.70$^{\rm c}$ &GTC & OSIRIS &\citet{estebanetal17}\\ 
Sh~2-212 & {\hii} & 14.6$\pm$1.4$^{\rm c}$ &GTC & OSIRIS &\citet{estebanetal17}\\ 
Sh~2-288 & {\hii} & 14.10$\pm$0.40$^{\rm c}$&GTC & OSIRIS &\citet{estebanetal17}\\ 
Sh~2-298 & WR-RN & 11.56$^{+0.87} _{-0.65}$$^{\rm b}$ &VLT & UVES&\citet{estebanetal17} \\
Sh~2-308 & WR-RN & 9.67$^{+0.31} _{-0.27}$$^{\rm b}$ &MTC & MagE &\citet{estebanetal16}\\ 
Sh~2-311 & {\hii} & 11.22 $\pm$ 0.22$^{\rm b}$&VLT & UVES &\citet{garciarojasetal05}\\ 
\hline
\end{tabular}
\begin{description} 
\item[$^{\rm a}$]{\hii}: {\hii} region; WR-RN: Wolf-Rayet ring nebula; O-RN: ring nebula around O-type star. 
\item[$^{\rm b}$]Distance determined from {\it Gaia} DR2 parallaxes (see text for details). 
\item[$^{\rm c}$]Distance taken from \citet{estebanetal17} or \citet{estebangarciarojas18}. 
\end{description} 
\end{table*}

\section{Results}\label{results}     

Although the set of source papers present calculations of physical conditions -- electron temperature, {\elect}, and density, {\elecd} -- and ionic and total abundances 
of several elements -- including helium in some cases-- for the nebulae, we decided to re-calculate all the relevant quantities in order to have a homogeneous set of values using the same methodology and updated atomic data. 

\begin{table*} 
\centering 
\caption{Atomic dataset used for collisionally excited lines of selected heavy-element ions.} 
\label{tab:atomic} 
\begin{tabular}{lcc} 
\hline 
& Transition probabilities &  \\ Ion & and energy levels & Collisional strengths \\ 
\hline 
N$^+$ & \citet{froesefischertachiev04} & \citet{tayal11} \\ 
O$^+$ & \citet{froesefischertachiev04} & \citet{kisieliusetal09} \\ 
O$^{2+}$ &  \citet{wieseetal96, storeyzeippen00} & \citet{storeyetal14} \\ 
S$^+$ & \citet{podobedovaetal09} & \citet{tayalzatsarinny10} \\ 
S$^{2+}$ &  \citet{podobedovaetal09} & \citet{grieveetal14} \\ 
\hline 
\end{tabular} 
\end{table*} 

 \subsection{Physical Conditions } 
 \label{conditions}
 
The values of {\elecd} and {\elect}  used for the calculation of the ionic abundances of the {\hii} regions have been taken from \citet{arellano-cordova20}. In the case of the ring nebulae -- G2.4$+$1.4, NGC~6888, NGC~7635, RCW52, RCW58, Sh~2-298 and Sh~2-308 --, we recalculate their {\elecd} and {\elect} using the same methodology and atomic data as for the rest of the objects. Following \citet{arellano-cordova20}, we use the version 1.0.26 of {\sc pyneb} \citep{luridianaetal15} in combination with the atomic data for collisionally excited lines listed in 
Table~\ref{tab:atomic} and the reddening corrected line-intensity ratios published in the source papers. The {\elecd} was calculated using the ratio  {\fsii} $\lambda6717/\lambda6731$ or its average with {\foii} $\lambda3729/\lambda3726$ when both density diagnostics were available. We obtain low densities 
 for most of the objects, and assumed the value {\elecd} = 100 {\cmc} \citep{osterbrockferland06} to determine {\elect} and ionic abundances when  {\elecd} 
 $<$ 100 {\cmc}. We used  a two-zone scheme characterized by 
{\elect}({\fnii}) and {\elect}({\foiii}) derived from the line intensity ratios {\fnii}~($\lambda6548+\lambda6584)/\lambda5755$ and {\foiii}~($\lambda4959+\lambda5007)/\lambda4363$, 
respectively. Although the intensity of the {\fnii} $\lambda$5755 auroral line may have some small contribution due to recombination,  we have not considered such effect in our calculations. We have estimated the contribution due to recombination using the formulae given by \citet{liuetal00} finding that it is almost negligible and always smaller than the uncertainties.  We have used the temperature relation proposed by \citet[their equation 3]{estebanetal09} -- calibrated with normal and giant {\hii} regions -- to estimate {\elect}({\fnii}) or {\elect}({\foiii}) in those objects where one of these temperature indicators was not available. 
We have used Monte Carlo calculations to estimate the uncertainties associated to each value of {\elecd} and {\elect}.  We generated 500 random values for each diagnostic line ratio 
assuming a Gaussian distribution with a standard deviation equal to the associated uncertainty of the line intensities involved in the diagnostic. With these distributions, we
calculated new simulated values of {\elecd} and {\elect}. Their associated errors correspond to a deviation of 68 percent -- equivalent to one standard
deviation -- centred in the mode of the distribution. The final results for {\elecd}, {\elect}({\fnii}) and {\elect}({\foiii}) and their associated uncertainties are included in Table~\ref{tab:ne_te_abund}. 

\begin{table*}
\centering
\caption{Physical conditions and some heavy-element ionic abundances$^{\rm a}$ for the sample objects.} 
\label{tab:ne_te_abund} 
\begin{tabular}{lccccccccc}
\hline
& & \elecd & \elect ({\foiii}) & \elect({\fnii}) & & & & & \\
Object& Zone & (cm$^{-3}$)& (K) & (K)& O$^{+}$& O$^{2+}$& S$^{+}$& S$^{2+}$& N$^{+}$ \\ \hline 
G2.4+1.4 & A1 &  110 $\pm$ 50& 8180 $\pm$ 1000$^{\rm b}$& 8780 $\pm$ 700& 8.10 $\pm$ 0.26& 8.64 $\pm$ 0.32& -& -& - \\
& A2 & 960 $\pm$ 160& 10370 $\pm$ 990& 8980 $\pm$ 350& 7.90 $\pm$ 0.12& 8.22 $\pm$ 0.22& 6.61 $\pm$ 0.07& 6.90 $\pm$ 0.13& 7.36 $\pm$ 0.05 \\
& A3 &  180 $\pm$ 70& 13520 $\pm$ 2990& 12510 $\pm$ 2090& 7.50 $\pm$ 0.46& 7.78 $\pm$ 0.47& -& -& - \\
& A4 & 100 $\pm$ 100& 10850 $\pm$ 1350& 10340 $\pm$ 1500& 7.49 $\pm$ 0.44& 8.22 $\pm$ 0.27& 6.00 $\pm$ 0.21& 6.42 $\pm$ 0.40& -  \\
& A5 &  110 $\pm$ 40& 9300 $\pm$ 1320$^{\rm b}$& 9560 $\pm$ 920& 7.96 $\pm$ 0.30& 8.43 $\pm$ 0.36& -& -& -  \\
M8 & & 1280 $\pm$ 480& 8040 $\pm$ 90& 8360 $\pm$ 110& 8.35 $\pm$ 0.06& 7.89 $\pm$ 0.03& 6.06 $\pm$ 0.04& 6.90 $\pm$ 0.02& -  \\
M16 & & 1120 $\pm$ 220& 7600 $\pm$ 210& 8320 $\pm$ 140& 8.46 $\pm$ 0.06& 7.92 $\pm$ 0.08& 6.44 $\pm$ 0.04& 6.86 $\pm$ 0.05& -  \\
M17 & & 470 $\pm$ 130& 7970 $\pm$ 120& 8900 $\pm$ 210& 7.79 $\pm$ 0.07& 8.45 $\pm$ 0.04& 5.53 $\pm$ 0.05& 6.93 $\pm$ 0.03& -  \\
M20 & & 260 $\pm$ 60& 7760 $\pm$ 250& 8260 $\pm$ 130& 8.48 $\pm$ 0.05& 7.74 $\pm$ 0.08& 6.31 $\pm$ 0.03& 6.83 $\pm$ 0.05& -  \\
M42 & & 6310 $\pm$ 2970& 8270 $\pm$ 50& 10140 $\pm$ 280& 7.74 $\pm$ 0.16& 8.44 $\pm$ 0.01& 5.48 $\pm$ 0.16& 6.85 $\pm$ 0.04& -  \\
NGC 2579  & & 950 $\pm$ 220& 9360 $\pm$ 170& 8640 $\pm$ 280$^{\rm c}$& 7.99 $\pm$ 0.09& 8.20 $\pm$ 0.05& 5.54 $\pm$ 0.06& 6.58 $\pm$ 0.03& -  \\
NGC 3576 & & 1490 $\pm$ 280& 8440 $\pm$ 60& 8760 $\pm$ 200& 8.04 $\pm$ 0.07& 8.37 $\pm$ 0.02& 5.79 $\pm$ 0.05& 6.88 $\pm$ 0.04& -  \\
NGC 3603 & & 2760 $\pm$ 910& 9020 $\pm$ 140& 11190 $\pm$ 550& 7.35 $\pm$ 0.13& 8.44 $\pm$ 0.04& 5.07 $\pm$ 0.10& 6.88 $\pm$ 0.04& -  \\
NGC 6888 & A1 &  100 $\pm$ 100& 19040 $\pm$ 1130& 7700 $\pm$ 700& 8.51 $\pm$ 0.28& 7.64 $\pm$ 0.05& 6.26 $\pm$ 0.11 & -& -  \\
& A2 & 250 $\pm$ 80& 12650 $\pm$ 410& 7410 $\pm$ 90& 8.31 $\pm$ 0.05& 7.53 $\pm$ 0.06& 6.27 $\pm$ 0.03& 6.09 $\pm$ 0.08& -  \\
& A3 & 200 $\pm$ 80& 9970 $\pm$ 660& 7580 $\pm$ 110& 8.13 $\pm$ 0.05& 7.71 $\pm$ 0.16& 6.15 $\pm$ 0.04& 6.38 $\pm$ 0.20& 7.74 $\pm$ 0.02 \\
& A4 & 120 $\pm$ 90& 10130 $\pm$ 970& 8870 $\pm$ 140& 7.46 $\pm$ 0.06& 7.93 $\pm$ 0.22& 5.53 $\pm$ 0.03& 6.58 $\pm$ 0.30& 8.06 $\pm$ 0.03 \\
& A5 & 170 $\pm$ 70& 9970 $\pm$ 550& 8740 $\pm$ 260& 7.47 $\pm$ 0.10& 7.95 $\pm$ 0.13& 5.55 $\pm$ 0.05& 6.68 $\pm$ 0.18& 8.01 $\pm$ 0.04 \\
& A6 & 170 $\pm$ 110& 9570 $\pm$ 780& 8720 $\pm$ 180& 7.54 $\pm$ 0.07& 8.00 $\pm$ 0.20& 5.54 $\pm$ 0.04& 6.74 $\pm$ 0.26& 7.33 $\pm$ 0.02 \\
NGC 7635 & A1 & 530 $\pm$ 90& 7240 $\pm$ 230$^{\rm b}$& 8110 $\pm$ 160& 8.32 $\pm$ 0.06& 7.92 $\pm$ 0.09& 6.42 $\pm$ 0.04& 7.05 $\pm$ 0.13& -  \\
& A2 & 120 $\pm$ 40& 8030 $\pm$ 520& 8040 $\pm$ 520& 7.98 $\pm$ 0.21& 8.29 $\pm$ 0.16& 5.74 $\pm$ 0.12& 7.06 $\pm$ 0.22& -  \\
& A3 & 100 $\pm$ 50& 7570 $\pm$ 610& 8200 $\pm$ 380& 8.08 $\pm$ 0.15& 8.33 $\pm$ 0.25& 5.84 $\pm$ 0.09& 7.16 $\pm$ 0.33& -  \\
& A4 & 1460 $\pm$ 150& 8710 $\pm$ 420& 9050 $\pm$ 90& 8.14 $\pm$ 0.03& 7.60 $\pm$ 0.11& 6.15 $\pm$ 0.03& 6.97 $\pm$ 0.15& -  \\
& A5 & 190 $\pm$ 100& 7280 $\pm$ 630$^{\rm b}$& 8140 $\pm$ 440& 8.20 $\pm$ 0.17& 8.17 $\pm$ 0.27& 6.04 $\pm$ 0.10& 6.96 $\pm$ 0.36& -  \\
& A6 & 100 $\pm$ 100& 7020 $\pm$ 710$^{\rm b}$& 7960 $\pm$ 500& 8.21 $\pm$ 0.20& 8.14 $\pm$ 0.31& 6.11 $\pm$ 0.10& 7.02 $\pm$ 0.43& -  \\
RCW 52 & & 230 $\pm$ 70& 5910 $\pm$ 620$^{\rm b}$& 7190 $\pm$ 430& 8.87 $\pm$ 0.23& 8.12 $\pm$ 0.34& 6.64 $\pm$ 0.14& 7.30 $\pm$ 0.20& -  \\
RCW 58 & & 100 $\pm$ 100& 4770 $\pm$ 250$^{\rm b}$& 6390 $\pm$ 180& 8.36 $\pm$ 0.11& 8.23 $\pm$ 0.23& 6.28 $\pm$ 0.06& 7.49 $\pm$ 0.14& 8.43 $\pm$ 0.05 \\
Sh~2-83 & & 300 $\pm$ 90& 10370 $\pm$ 370& 11960 $\pm$ 640& 7.15 $\pm$ 0.12& 8.25 $\pm$ 0.07& 5.20 $\pm$ 0.07& 6.51 $\pm$ 0.10& -  \\
Sh~2-100 & & 420 $\pm$ 240& 8250 $\pm$ 150& 8610 $\pm$ 250& 7.73 $\pm$ 0.10& 8.41 $\pm$ 0.04& 5.53 $\pm$ 0.07& 6.96 $\pm$ 0.06& -  \\
Sh~2-127 & & 600 $\pm$ 100& 9660 $\pm$ 220$^{\rm b}$& 9810 $\pm$ 150& 8.20 $\pm$ 0.05& 7.88 $\pm$ 0.21& 5.91 $\pm$ 0.03& 6.72 $\pm$ 0.07& -  \\
Sh~2-128 & & 480 $\pm$ 100& 9970 $\pm$ 340& 10550 $\pm$ 240& 7.84 $\pm$ 0.06& 7.98 $\pm$ 0.07& 5.62 $\pm$ 0.04& 6.56 $\pm$ 0.09& -  \\
Sh~2-152 & & 750 $\pm$ 80& 7360 $\pm$ 120& 8200 $\pm$ 70& 8.46 $\pm$ 0.03& 7.70 $\pm$ 0.05& 6.15 $\pm$ 0.02& 7.18 $\pm$ 0.06& -  \\
Sh~2-209 & & 300 $\pm$ 290& 10760 $\pm$ 1140$^{\rm b}$& 10580 $\pm$ 800& 7.67 $\pm$ 0.31& 7.88 $\pm$ 0.21& 5.51 $\pm$ 0.11& 6.29 $\pm$ 0.29& -  \\
Sh~2-212 & & 100 $\pm$ 100& 11250 $\pm$ 970& 8350 $\pm$ 770& 8.16 $\pm$ 0.31& 7.81 $\pm$ 0.17& 5.16 $\pm$ 0.18& 5.94 $\pm$ 0.23& -  \\
Sh~2-288 & & 410 $\pm$ 270& 9200 $\pm$ 530& 9430 $\pm$ 340& 8.20 $\pm$ 0.11& 7.75 $\pm$ 0.15& 5.93 $\pm$ 0.07& 6.67 $\pm$ 0.19& -  \\
Sh~2-298 & & 100 $\pm$ 100& 11720 $\pm$ 210& 11650 $\pm$ 490& 8.15 $\pm$ 0.10& 8.10 $\pm$ 0.04& 6.48 $\pm$ 0.06& 6.50 $\pm$ 0.03& 7.32 $\pm$ 0.04 \\
Sh~2-308 & & 100 $\pm$ 100& 16600 $\pm$ 2470& 14300 $\pm$ 2910& 6.92 $\pm$ 0.42& 7.84 $\pm$ 0.24& 5.32 $\pm$ 0.26& 6.53 $\pm$ 0.16& 7.08 $\pm$ 0.18 \\
Sh~2-311 & & 290 $\pm$ 80& 8940 $\pm$ 110& 9270 $\pm$ 180& 8.28 $\pm$ 0.06& 7.83 $\pm$ 0.03& 6.21 $\pm$ 0.04& 6.70 $\pm$ 0.03& -  \\\hline
\end{tabular}
\begin{description} 
\item[$^{\rm a}$] In units of 12 + log(\ionic{X}{n+}/\ionic{H}{+}). 
\item[$^{\rm b}$] \elect ({\foiii}) estimated from {\elect}({\fnii}) using equation 3 of \citet{estebanetal09}.
\item[$^{\rm c}$] \elect ({\fnii}) estimated from {\elect}({\foiii}) using equation 3 of \citet{estebanetal09}.
\end{description} 
\end{table*} 

\subsection{He$^+$  abundances} 
\label{ionic}

All the spectra of our sample show several {\hei} recombination lines.  Although the set of source papers present calculations of the 
\ionic{He}{+}/\ionic{H}{+}  and He/H ratios of some of the nebulae, we have re-calculated them using the physical conditions indicated in Section~\ref{conditions}. For each object, we use all or 
several of the following list of {\hei} lines: \lamb3614, \lamb3889, \lamb3965, \lamb4026, \lamb4121, \lamb4388, \lamb4438, \lamb4471, \lamb4713, \lamb4922, \lamb5016, \lamb5048, \lamb5876, \lamb6678, 
\lamb7065, \lamb7281 and \lamb9464. The He$^{+}$ abundance has been determined using {\sc pyneb}, the aforementioned line intensity ratios taken from the set of source papers, the values of {\elecd} and {\elect}({\foiii})  
given in Table\ref{tab:ne_te_abund} and the effective recombination coefficient computations by \citet{porteretal12, porteretal13} for {\hei} lines and \citet{storeyhummer95} for {\hi} lines. 
The calculations by \citet{porteretal12} include collisional effects in the level populations of \ionic{He}{0}. A Monte Carlo simulation similar to the one described in Section~\ref{conditions} 
-- including random distributions of {\elecd}, {\elect} and the line intensities -- was applied to estimate the uncertainty of the \ionic{He}{+}/\ionic{H}{+} ratios derived for each individual line of each spectrum. The \ionic{He}{+}/\ionic{H}{+} ratios obtained for each of the {\hei} lines of the spectra of the sample objects are shown in Table~\ref{tab:hep}. Lines with intensity uncertainties greater than 40\% and those affected by blending with telluric lines or other spectral features were not considered for abundance determinations. Moreover, the \ionic{He}{+}/\ionic{H}{+} ratios calculated from the bright {\hei} $\lambda$3889 and 
$\lambda$7065 lines are not included in Table~\ref{tab:hep}. Both lines are the most affected, with great difference, by optical depth effects of the 2~$^3S$ level on the predicted recombination spectrum of {\hei} \citep{benjaminetal02}, that only affect noticeably to triplet lines. Moreover, {\hei} $\lambda$3888.64 is blended with the also bright {\hi} $\lambda$3889.05 line even in our highest spectral resolution spectra. 

We have calculated the weighted mean of the individual  \ionic{He}{+}/\ionic{H}{+} ratios of each spectrum. The weight of each line has 
been taken as the inverse of the square of the error associated with its \ionic{He}{+}/\ionic{H}{+} value. For a given spectrum, we have calculated means taking all possible combinations of the individual lines without repetition, leaving at least 3 lines to average (for those spectra with more than 3 {\hei} lines). Once this is done, of all the means, we consider those within the 5th percentile of uncertainty and, from this subset, we finally select the mean for which more {\hei} lines have been used for its calculation. This method excludes the most discrepant individual lines but maintain the largest possible number of them giving consistent values. The finally adopted mean \ionic{He}{+}/\ionic{H}{+} ratio of each spectrum is also included in Table~\ref{tab:hep}. The uncertainty associated to that mean corresponds to the weighted standard deviation. In Table~\ref{tab:list}, we give the list of individual {\hei} lines used to derive the average value of the \ionic{He}{+}/\ionic{H}{+} ratio adopted for each spectrum. 

The usual methodology for calculating chemical abundances in H~{\sc ii} regions is to consider a zone of high and low ionization and adopt a representative {\elect} for each zone. However, this may be inappropriate for several ions, because they can emit radiation in an intermediate zone or in both ionization zones. This is the case of \ionic{He}{+}, since it can emit in the high and low ionization zone. To calculate \ionic{He}{+} optimally, the characteristic temperature of the ion -- {\elect}({\hei}) in this case -- should be used or the precise geometry of the different ionization zones of each nebula should be known. However, the difference of \ionic{He}{+}/\ionic{H}{+}  considering {\elect}({\hei}) and {\elect}({\foiii}) is expected to be small given the small dependence on temperature of the intensity ratios of recombination lines. The difference between {\elect}({\hei}) and {\elect}({\foiii}) is important for those interested in determining the primordial helium abundance, with the least possible uncertainty ($\leq$ 1\%). In our case, an uncertainty of the order or even somewhat larger than 1\% in the calculation of the \ionic{He}{+}/\ionic{H}{+} ratio is acceptable, even more taking into account that statistically those uncertainties will not be reflected in the determination of a gradient based on data of several objects and whose main source of uncertainty is not  \ionic{He}{+}/\ionic{H}{+} but the \ionic{He}{0}/\ionic{H}{+} ratio. Nevertheless, we have explored the effects that {\elect} and optical depth effects of the metastable 2~$^3S$ level  -- $\tau(2^3S)$ but  parametrized by $\tau$(3889) -- would have on the \ionic{He}{+}/\ionic{H}{+} ratios.

\begin{figure} 
\centering 
\includegraphics[scale=0.335]{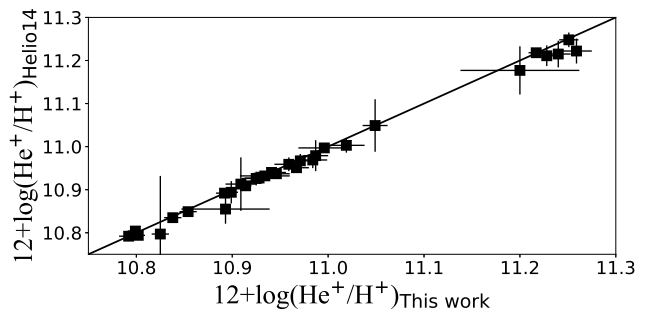} 
 \caption{General comparison of the average \ionic{He}{+}/\ionic{H}{+} ratios obtained for the sample objects using our methodology described in the text and using the \texttt{Helio14} code \citep{Peimbert2012}. The continuous line indicates the 1:1 relation.} 
 \label{fig:comp_gral_helio14} 
 \end{figure}

Figure~\ref{fig:comp_gral_helio14} presents the comparison between the \ionic{He}{+}/\ionic{H}{+} ratios obtained from the method followed in this work and those obtained using \texttt{Helio14} with the list of lines presented in Table~\ref{tab:list} and {\elect}(\oii+\oiii), derived  following \citet{peimbertetal2002}. {\elect}(\oii+\oiii) takes into account the emission of {\hei} in the low ionization zone. \texttt{Helio14} code is an updated version of  \texttt{Helio10} presented by \citet{Peimbert2012}. The code uses the effective recombination coefficients of \citet{storeyhummer95} for {\hi}, and \citet{Porter2007} for {\hei}, the collisional contribution of {\hei} lines calculated by \citet{saweyberrington93} and the optical depth effects in the triplets estimated by \citet{kingdonferland95}. \texttt{Helio14} calculates the most likely values for {\elecd}({\hei}), $\tau(2^3S)$ and the  \ionic{He}{+}/\ionic{H}{+} ratio from the theoretical {\hei}/H$\beta$ ratios using as input a set of parameters, atomic data, and up to 20 {\hei}/H$\beta$ line intensity ratios along with their uncertainties. Then, it compares the observed ratios to the theoretical ones minimising $\chi^2$:

\begin{equation}
\chi^2=\sum_\lambda\frac{\left \{ \left[\frac{I(\lambda)}{I({\rm H}\beta)}\right]_{\rm obs}-\left[\frac{I(\lambda)}{I({\rm H}\beta)}\right]_{\rm theo}\right\} ^2}{\sigma_I(\lambda)^2}.
\end{equation}

Where $\sigma_I(\lambda)$ is the uncertainty in the intensity of \lamb. The comparison of Figure~\ref{fig:comp_gral_helio14} shows that all the objects are entirely consistent with the 1:1 relation. The nebulae with the largest deviations with respect the 1:1 line are RCW 52, Sh~2-209 and Sh~2-308, although the values agree with such relation  within the uncertainties. These three objects are among the ones with fewer {\hei} lines in their  spectra: 4, 4 and 2, respectively, so the determination of the abundance of {\hei} is very sensitive to the variations of the fitted parameters in the minimisation of $\chi^2$.  \texttt{Helio14} determines the available parameter space within $\chi^2_{min}<\chi^2<\chi^2_{min}+1$ to estimate the 1$\sigma$ error bars. In any case, from these 3 objects, just RWC 52 was taken into account in the determination of radial {\hei} gradients. The rest are ring nebula associated Wolf-Rayet stars and they present an overabundance of helium. This is discussed in greater detail in \S\ref{sec:HeWR}. The general deviation between the values of \ionic{He}{+}/\ionic{H}{+} ratios obtained with \texttt{Helio14} and those obtained following the procedure described in this section is less than 0.008 dex even including the 3 objects showing the largest differences.

From all the spectra in our sample, we have deconvolved {\hei} \lamb3889.64 from {\hi} \lamb3889.05 by multiple Gaussian fitting in NGC 3576 and Sh~2-311. Among all the spectra, only Sh~2-152, NGC 3576, NGC 7635-A4, M16, M17, M20 and M42 have measurements of the intensity ratio of {\hei} \lamb3889.64 or the blend of {\hei} \lamb3889.64 and {\hi} \lamb3889.05 with  H$\beta$ with uncertainties smaller than 3\%. In these last cases, we study the possible effect of $\tau$(3889) on the determination of the  \ionic{He}{+}/\ionic{H}{+} ratios. A small uncertainty in the measurement of  {\hei} \lamb3889.64 is indispensable to avoid  spreading uncontrolled errors in $\tau$(3889) and consequently in the abundance of \ionic{He}{+}/\ionic{H}{+}. 

\begin{figure} 
\centering 
\includegraphics[scale=0.335]{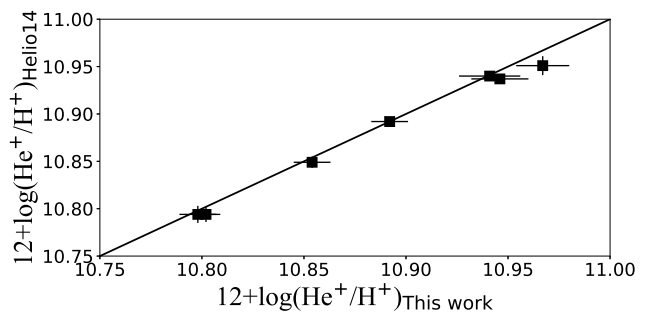} 
 \caption{Comparison between \ionic{He}{+}/\ionic{H}{+} determinations after estimating the effects of optical depth ($\tau$(3889)) in 7 objects: Sh~2-152, NGC 3576, NGC 7635-A4, M16, M17, M20 and M42. The continuous line indicates the 1:1 relation.} 
 \label{fig:comp_3889_helio14} 
 \end{figure}

In the nebulae where both lines cannot be deblended with multiple Gaussian fitting, the contribution of {\hei} $\lambda$3889.64 to the total intensity was obtained after subtracting its theoretical ratio to {\hi} $\lambda$3889.05 based on the work by \citet{Brocklehurst1971}. In Figure~\ref{fig:comp_3889_helio14} we compare the values of the \ionic{He}{+}/\ionic{H}{+} ratios for Sh~2-152, NGC 3576, NGC 7635-A4, M16, M17, M20 and M42 using \texttt{Helio14} with the same input parameters as in Figure~\ref{fig:comp_gral_helio14} but taking into account $\tau$(3889) and the results we obtain with our methodology. The general dispersion is smaller than 0.007 dex. From this comparison, we can conclude that our results can be considered virtually independent of the effects of $\tau(2^3S)$. In fact, this is minimised by averaging \ionic{He}{+}/\ionic{H}{+} ratios from a large number of emission lines, including several singlets, which intensities are practically independent of $\tau(2^3S)$.

Among the brightest {\hei} lines, \lamb4922, \lamb5876 and \lamb6678 have been considered for calculating the weighted mean in more that 84\% of the spectra in which they have been measured (most or all of them). Other well-behaved less intense {\hei} lines are 
\lamb3614, \lamb3965, \lamb4026,  \lamb4388 and \lamb4438, that have been taken for calculating the mean in more than 70\% of the spectra in which they 
have been measured. Most of these lines (6) corresponds to singlets, and only 2 to triplets (\lamb4026 and \lamb5876). The {\hei} lines that have a greater tendency to provide less consistent values with respect to the average \ionic{He}{+}/\ionic{H}{+} ratio are: \lamb4121, \lamb4713, \lamb5048, \lamb7281 and \lamb9464, that have been used in less than 50\% of the objects in which they have been measured. Of these, 2 are singlets and 3 triplets. In the case of the 2 lines of redder wavelength, \lamb7281 and \lamb9464, their line intensity may be affected by contamination with telluric emission or absorption features.  In fact, when these lines are observed with high-resolution echelle spectroscopy, the \ionic{He}{+} abundances obtained from them are similar to those obtained from other more intense lines and they are considered to compute the average value of the  \ionic{He}{+}/\ionic{H}{+} ratio. The {\hei} \lamb5015 line is relatively bright but has not been considered in many spectra because it is blended with the {\foiii} $\lambda$5007 line in the OSIRIS spectra and affected by a ghost emission feature in the UVES ones. Another bright {\hei} line is $\lambda$4471, but it is only used in 59\% of the spectra. This is mainly because it provides \ionic{He}{+}/\ionic{H}{+} ratios lower than the mean in G2.4+1.4, NGC~6888 and NGC~7635, the objects with the largest number of spectra. The whole optical OSIRIS spectra of these objects were taken with two grisms, R1000B and R2500V. Since the {\hei} $\lambda$4471 line was the closest one to the blue edge of the spectra taken with in R2500V, some deviations at the border of the flux calibration curve may be affecting the flux measurement obtained by \citet{estebanetal16} for this line. 

\subsection{Total He abundances} \label{totalHe}

The presence of \ionic{He}{0} inside the \ionic{H}{+} zone may be significant in low ionization nebulae. We need to assume an ICF(He) to determine the total helium 
abundance from the measured \ionic{He}{+}/\ionic{H}{+} ratio, in the form: He/H = ICF(He) $\cdot$ \ionic{He}{+}/\ionic{H}{+}. However, the estimation of the ICF(He) is a controversial issue and there is no a consensus among the different authors. Traditional ICF(He) schemes involve abundance ratios of different heavy-element ions such as \ionic{O}{+}, \ionic{O}{2+}, \ionic{S}{+} or \ionic{S}{2+}. 
For the {\hii} regions of our sample, we have used the values of \ionic{O}{+}, \ionic{O}{2+}, \ionic{S}{+} and \ionic{S}{2+} abundances determined by \citet{arellano-cordova20}, who use {\sc pyneb} and the atomic data given in Table~\ref{tab:atomic}. In addition, we have re-calculated the \ionic{N}{+} abundances of the WR ring nebulae using the same methodology, their N abundances will be discussed in \S\ref{sec:HeWR}. The line intensities used to derive the heavy-element ionic abundances have been {\foii} $\lambda\lambda3726, 3729$, {\foii} $\lambda\lambda4959, 5007$, {\fsii} $\lambda\lambda6717, 6731$, {\fsiii} $\lambda$6312 (we used also {\fsiii} $\lambda\lambda$ 9069, 9035 in several objects) and {\fnii} $\lambda\lambda6548, 6583$. Their reddening-corrected values have been taken from the set of source papers. \citet{arellano-cordova20} use {\elect}({\fnii}) as representative of the O$^{+}$, S$^{+}$ and N$^{+}$ emitting regions and {\elect}({\foiii}) as representative of the O$^{++}$ zone. Following the recommendation of \citet{dominguezguzmanetal19}, \citet{arellano-cordova20} adopt the mean value of {\elect}({\fnii}) and {\elect}({\foiii}) to calculate the S$^{++}$ abundance. In the sake of homogeneity and as we did for determining the physical conditions, we used the same methodology and atomic data as \citet{arellano-cordova20} for deriving the \ionic{O}{+}, \ionic{O}{2+}, \ionic{S}{+}, \ionic{S}{2+} or \ionic{N}{+}abundances for the ring nebulae. The heavy-element ionic abundances determined for each spectrum and object are included in Table~\ref{tab:ne_te_abund}.

\begin{table*}
\centering
\caption{Total He abundances$^{\rm a}$ for the sample objects using different ICF schemes.} 
\label{tab:he} 
\begin{tabular}{lcccc}
\hline
& \multicolumn{4}{c}{ICF(He) scheme$^{\rm b}$} \\
Object & PTP77 & ZL03 & KS83 & B09   \\
\hline
G2.4+1.4  &  \multicolumn{4}{c}{11.038 $\pm$ 0.017$^{\rm c}$} \\
M8 &  10.963 $\pm$ 0.038& 10.897 $\pm$ 0.016 & 10.927 $\pm$ 0.032 & 11.098 $\pm$ 0.093\\
M16 &  11.094 $\pm$ 0.066& 11.031 $\pm$ 0.033 & 10.985 $\pm$ 0.034 & 10.982 $\pm$ 0.106\\
M17 &  10.994 $\pm$ 0.020& 10.984 $\pm$ 0.016 & 10.987 $\pm$ 0.018 & 10.987 $\pm$ 0.052\\
M20 &  11.043 $\pm$ 0.055& 10.968 $\pm$ 0.028 & 10.957 $\pm$ 0.033 & 11.095 $\pm$ 0.125\\
M42 &  10.981 $\pm$ 0.031& 10.964 $\pm$ 0.022 & 10.964 $\pm$ 0.022 & 10.953 $\pm$ 0.054\\
NGC 2579 &  10.998 $\pm$ 0.046& 10.972 $\pm$ 0.033 & 10.978 $\pm$ 0.040 & 10.976 $\pm$ 0.095\\
NGC 3576 &  10.995 $\pm$ 0.028& 10.975 $\pm$ 0.021 & 10.977 $\pm$ 0.023 & 10.964 $\pm$ 0.054\\
NGC 3603 &  11.008 $\pm$ 0.032& 11.002 $\pm$ 0.029 & 11.004 $\pm$ 0.030 & 11.010 $\pm$ 0.082\\
NGC 6888 &   \multicolumn{4}{c}{11.244 $\pm$ 0.011$^{\rm c}$} \\
NGC 7635 & 11.051 $\pm$ 0.030& 11.031 $\pm$ 0.034 & 11.041 $\pm$ 0.034 & 11.087 $\pm$ 0.057\\
RCW 52 &  10.960 $\pm$ 0.162& 10.903 $\pm$ 0.070 & 10.921 $\pm$ 0.134 & 11.144 $\pm$ 0.514\\
RCW 58 &  11.279 $\pm$ 0.046& 11.242 $\pm$ 0.019 & 11.284 $\pm$ 0.051 & 11.503 $\pm$ 0.306\\
Sh~2-83 &  10.916 $\pm$ 0.015& 10.921 $\pm$ 0.014 & 10.908 $\pm$ 0.010 & 10.997 $\pm$ 0.163\\
Sh~2-100 & 11.009 $\pm$ 0.024& 11.000 $\pm$ 0.020 & 11.003 $\pm$ 0.021 & 11.003 $\pm$ 0.071\\
Sh~2-127 &  10.944 $\pm$ 0.045& 10.863 $\pm$ 0.022 & 10.906 $\pm$ 0.034 & 10.967 $\pm$ 0.195\\
Sh~2-128 &  11.004 $\pm$ 0.040& 10.976 $\pm$ 0.029 & 10.976 $\pm$ 0.029 & 10.957 $\pm$ 0.083\\
Sh~2-152 &  10.920 $\pm$ 0.028& 10.840 $\pm$ 0.014 & 10.906 $\pm$ 0.024 & 11.327 $\pm$ 0.090\\
Sh~2-209 &  10.979 $\pm$ 0.154& 10.960 $\pm$ 0.104 & 10.936 $\pm$ 0.100 & 10.896 $\pm$ 0.081\\
Sh~2-212 &  11.128 $\pm$ 0.155& 11.085 $\pm$ 0.077 & 11.101 $\pm$ 0.128 & 11.197 $\pm$ 0.482\\
Sh~2-288 & 10.928 $\pm$ 0.090& 10.864 $\pm$ 0.050 & 10.880 $\pm$ 0.052 & 10.998 $\pm$ 0.289\\
Sh~2-298 &  11.252 $\pm$ 0.138& 11.260 $\pm$ 0.062 & 11.032 $\pm$ 0.040 & 11.136 $\pm$ 0.134\\
Sh~2-308 &  11.222 $\pm$ 0.094& 11.226 $\pm$ 0.086 & 11.212 $\pm$ 0.080 & 11.264 $\pm$ 0.469\\
Sh~2-311 &  11.080 $\pm$ 0.046& 11.037 $\pm$ 0.021 & 11.002 $\pm$ 0.030 & 10.992 $\pm$ 0.070\\
\hline
\end{tabular}
\begin{description} 
\item[$^{\rm a}$] In units of 12 + log(He/H). 
\item[$^{\rm b}$] PTP77: \citet{peimberttorrespeimbert77}; ZL03: \citet{zhangliu03}; KS83: \citet{kunthsargent83}; B09: \citet{bresolinetal09}.
\item[$^{\rm c}$] He/H = \ionic{He}{+}/ \ionic{H}{+} + \ionic{He}{2+}/ \ionic{H}{+} for this object. No ICF(He) applied. 
\end{description} 
\end{table*} 

Photoionization models \citep[e.g.][]{stasinskaschaerer97} predict that helium is completely ionized inside the  \ionic{H}{+} zone when the
exciting star is hotter than about 39 000K (earlier than O6.5V), independently of the ionization parameter. Most of the ionizing stars of our sample are ionized by 
stars that should be colder, so one would expect a certain amount of  \ionic{He}{0} in most of our nebulae. On the other hand, observations in different directions inside {\hii} regions carried out by \citet{deharvengetal00} show that even an  {\hii} region excited by a star of spectral type earlier than O6.5 can contain a significant amount
of  \ionic{He}{0}. We have used different ICF(He) schemes because, as we have said before,  
there is not a standard method to correct for the fraction of  \ionic{He}{0}, although all schemes consider that the higher the ionization degree the higher the \ionic{He}{+}/He ratio of the nebulae. ICF(He) schemes usually consider the similarity between the ionization potential of  \ionic{He}{0} (24.59 eV) and those of 
S$^{+}$ (23.33 eV) and/or O$^{+}$ (34.97 eV) assuming that the relation \ionic{O}{+}/O~$>$~\ionic{He}{0}/He~$>$\ionic{S}{+}/S should be applicable but using different parameters proportional to the ionization degree of the nebula. One of the first ICF(He) schemes was proposed by \citet{peimberttorrespeimbert77} in their work on the chemical composition of the Orion Nebula that uses a linear combination of 
\ionic{O}{+}/O and \ionic{S}{+}/S ratios, 
\begin{equation}
    \mathrm{ICF(He)_{PTP77}} = \left[1 - \gamma \cdot \frac{\mathrm{O}^+}{\mathrm{O}} - (1 - \gamma) \cdot \frac{\mathrm{S}^+}{\mathrm{S}}\right]^{-1}. 
	\label{eq:PTP77}
\end{equation}
The value of $\gamma$ depends on the density distribution \citep{peimbertetal74}. It can be determined when spectra at different positions covering different ionization conditions are available for the nebula, but this is not the case for most of our sample objects. The only objects with several slit positions are all ring nebulae and 
bubbles, that are better represented by relatively thin ionized shells rather than typical Strömgren spheres and, therefore, that method for estimating the parameter $\gamma$ may not be appropriate. \citet{peimberttorrespeimbert77} estimated $\gamma$ = 0.35 in the case of the Orion Nebula and \citet{peimbertetal78} found $\gamma$ = 0.20 for $\eta$ Carina nebula. Those last authors assume the same value of 0.20 for other {\hii} regions with {\elecd} values similar to that of  $\eta$ Carina nebula (300-1000 {\cmc}). \citet{lequeuxetal79} use equation~\ref{eq:PTP77} for determining the He abundance for a sample of metal-poor {\hii} galaxies with {\elecd} in the 10-100 {\cmc} range using $\gamma$ = 0.15. In our case, following the prescriptions given in the cited works, we have used equation~\ref{eq:PTP77} using values of $\gamma$ interpolating between 0.15 to 0.35 as a function of the {\elecd} of the nebulae. 

Other ICF(He) schemes only use \ionic{O}{+}/O or \ionic{S}{+}/S ratios as indicators of the ionization degree of the nebulae. \citet{kunthsargent83}, in their study of a sample of metal-poor galaxies, proposed an ICF(He) dependent on the \ionic{O}{+}/O ratio deduced from \citet{peimbertetal74} empirical models, 
\begin{equation}
    \mathrm{ICF(He)_{KS83}} = \left[1 - 0.25 \cdot \frac{\mathrm{O}^+}{\mathrm{O}}\right]^{-1}. 
	\label{eq:KS83}
\end{equation}
 \citet{peimbertetal92} presented a detailed spectroscopical study of the highly-ionized Galactic {\hii} region M17 and propose an ICF(He) parameterized only by the \ionic{S}{+}/S ionization fraction.  \citet{zhangliu03} use a very similar scheme in their study of the relatively low excitation planetary nebula M~2-24 but depending on the \ionic{S}{+}/\ionic{S}{2+} ratio, which -- in contrast to \ionic{S}{+}/S -- is a parameter that can be obtained directly from optical spectra: 
\begin{equation}
    \mathrm{ICF(He)_{ZL03}} =  1 + \frac{\mathrm{S}^+}{\mathrm{S}^{2+}}. 
	\label{eq:ZL03}
\end{equation}

\citet{delgadoingladaetal14} do not recommend the use of traditional ICF(He) schemes -- as those given in equations \ref{eq:KS83} and \ref{eq:ZL03} -- because the \ionic{He}{0}/\ionic{He}{+} ratio is more dependent on the effective temperature of the ionizing source, whereas heavy-element abundance ratios depend essentially on the ionization parameter. 
\citet{vilchez89} proposed an ICF(He) scheme based on the radiation softness parameter defined as $\eta$ = (\ionic{O}{+}/\ionic{S}{+}) $\cdot$ (\ionic{S}{2+}/\ionic{O}{2+}) \citep{vilchezpagel88}, which is sensitive to the effective temperature of the ionizing source and a good indicator of the ionization structure of the nebula. 
\citet{vilchez89} and \citet{pageletal92} pointed out that for  log($\eta$) $<$ 0.9 the amount of  \ionic{He}{0} inside the \ionic{H}{+} zone is negligible for a large variety of models, but becomes very model dependent for  log($\eta$) $>$ 1. \citet{izotovetal94} and \citet{bresolinetal09} have obtained approximate analytical expressions of ICF(He) as a function of $\eta$ parameter based on photoionization models. In our case, we have used the relation given by \citet{bresolinetal09} in their study of {\hii} regions in the spiral galaxy NGC~300 and based on the predictions of photoionization models by \citet{stasinskaetal01},
\begin{equation}
    \mathrm{ICF(He)_{B09}} = 1.585 + \log( \eta) \cdot [1.642 \cdot \log(\eta) - 1.948]. 
	\label{eq:B09}
\end{equation}

In the case of the ring nebulae for which we have spectra at several slit positions and, therefore, several independent determinations of the He abundance, we have selected the representative He/H ratio 
of the nebula as a whole attending to different considerations in each object. In the case of G2.4$+$1.4, we only considered the spectra of zone A2, which has the highest signal-to-noise ratio and was observed with higher spectral resolution (MagE echelle spectrograph). In the case of NGC~6888, we take the weighted mean value of zones A3, A4, A5 and A6, which are located in the main body of the ionized nebula. Zones A1 and A2 correspond to high-ionization narrow peripheral arcs showing very high values of {\elect} and may be contaminated by shock excitation \citep[see][]{estebanetal16}. For NGC~7635 we have taken the weighted mean value of zones A2, A3, A5 and A6, which correspond to the relatively highly-ionized gas of the expanding bubble. Zone A1 of NGC~7635 corresponds to one of the brightest knots of Sh 2-162, a large {\hii} region that encompasses the ring nebula NGC 7635. Sh 2-162 is ionized by the same O-type star but shows a lower ionization degree. Zone A4 covers the brightest of a group of several high-density cometary knots just at the southwest of the ionizing star that show rather higher densities and quite lower ionization degrees. 

The WR ring nebulae G2.4$+$1.4 and NGC~6888 show {\heii} lines in their spectra \citep{estebanetal16}. G2.4$+$1.4 is a high ionization degree object ionized by a very hot WO2 star and the {\heii} \lamb4686 line is very bright. We have determined the  \ionic{He}{2+} abundance 
using {\sc pyneb}, {\elect}({\foiii}) and the effective recombination coefficient calculated by \citet{storeyhummer95}, obtaining 12 $+$ log(\ionic{He}{2+}/\ionic{H}{+}) = 10.45 $\pm$ 0.02 for the zone A2 of G2.4$+$1.4. In the case on NGC~6888 the \ionic{He}{2+} abundances we determine for the different zones is very much lower, with values ranging from 8.14 to 8.51, which are clearly negligible in comparison to the \ionic{He}{+}/\ionic{H}{+} ratio determined in this object. 
We did not use an ICF(He) for G2.4$+$1.4 and NGC~6888  and their total He abundance was simply  \ionic{He}{+}/ \ionic{H}{+} + \ionic{He}{2+}/ \ionic{H}{+}. 

In Table~\ref{tab:he} we show the He/H ratios obtained for each nebulae using the different ICF(He) schemes considered. In the case of the 
ICF(He)$_\mathrm{PTP77}$, there is an explicit dependence of the S/H ratio, which has been derived from the sum of the \ionic{S}{+} and  \ionic{S}{2+} abundances given in  Table~\ref{tab:ne_te_abund} and an appropriate ICF(S) using the relations: 
\begin{equation}
    \frac{\mathrm{S}}{\mathrm{H}} =  \mathrm{ICF(S)} \cdot \left[ \frac{\mathrm{S}^+}{\mathrm{H}^{+}} + \frac{\mathrm{S}^{2+}}{\mathrm{H}^{+}} \right], 
	\label{eq:abS}
\end{equation}
where 
\begin{equation}
    \mathrm{ICF(S)} = \left[1 - \left(1 - \frac{\mathrm{O}^+}{\mathrm{O}} \right)^3 \right]^{-1/3}, 
	\label{eq:ICFS}
\end{equation}
which was proposed by \citet{stasinska78}. The uncertainty of the total He abundance has been determined propagating the errors of the  \ionic{He}{+}/\ionic{H}{+} ratio and that of the ICF(He) -- and ICF(S) -- used, which comes from the derivation of each equation of the ICF and the quoted errors of the ionic abundances given in Table~\ref{tab:ne_te_abund}.

\begin{table*}
\centering
\caption{Radial He/H gradients using different combinations of {\hii} regions and ICF schemes.} 
\label{tab:gradHe} 
\begin{tabular}{lcccc}
\hline
& \multicolumn{2}{c}{All {\hii} regions} & \multicolumn{2}{c}{Only with log($\eta$) $<$ 0.9} \\
ICF(He) & Slope (dex kpc$^{-1})$ & Intercept & Slope (dex kpc$^{-1})$ & Intercept \\
\hline
PTP77 & $-$0.0036 $\pm$ 0.0069 & 11.04 $\pm$ 0.07 & $-$0.0046 $\pm$ 0.0091 & 11.04 $\pm$ 0.08 \\
ZL03 & $-$0.0028 $\pm$ 0.0043 & 10.99 $\pm$ 0.04 & $-$0.0044 $\pm$ 0.0062 & 11.02 $\pm$ 0.06 \\
KS83 & $-$0.0025 $\pm$ 0.0049 & 10.99 $\pm$ 0.05 & $-$0.0067 $\pm$ 0.0060 & 11.04 $\pm$ 0.06 \\
B09 &  $-$0.006 $\pm$ 0.015 & 11.09 $\pm$ 0.07 & $-$0.0051 $\pm$ 0.0077 & 11.09 $\pm$ 0.15 \\
No ICF(He) & $--$ & $--$ & $-$0.0078$\pm$ 0.0029 & 11.03 $\pm$ 0.03 \\
\hline
\end{tabular}
\end{table*} 

\begin{figure*} 
\centering 
\includegraphics[scale=0.17]{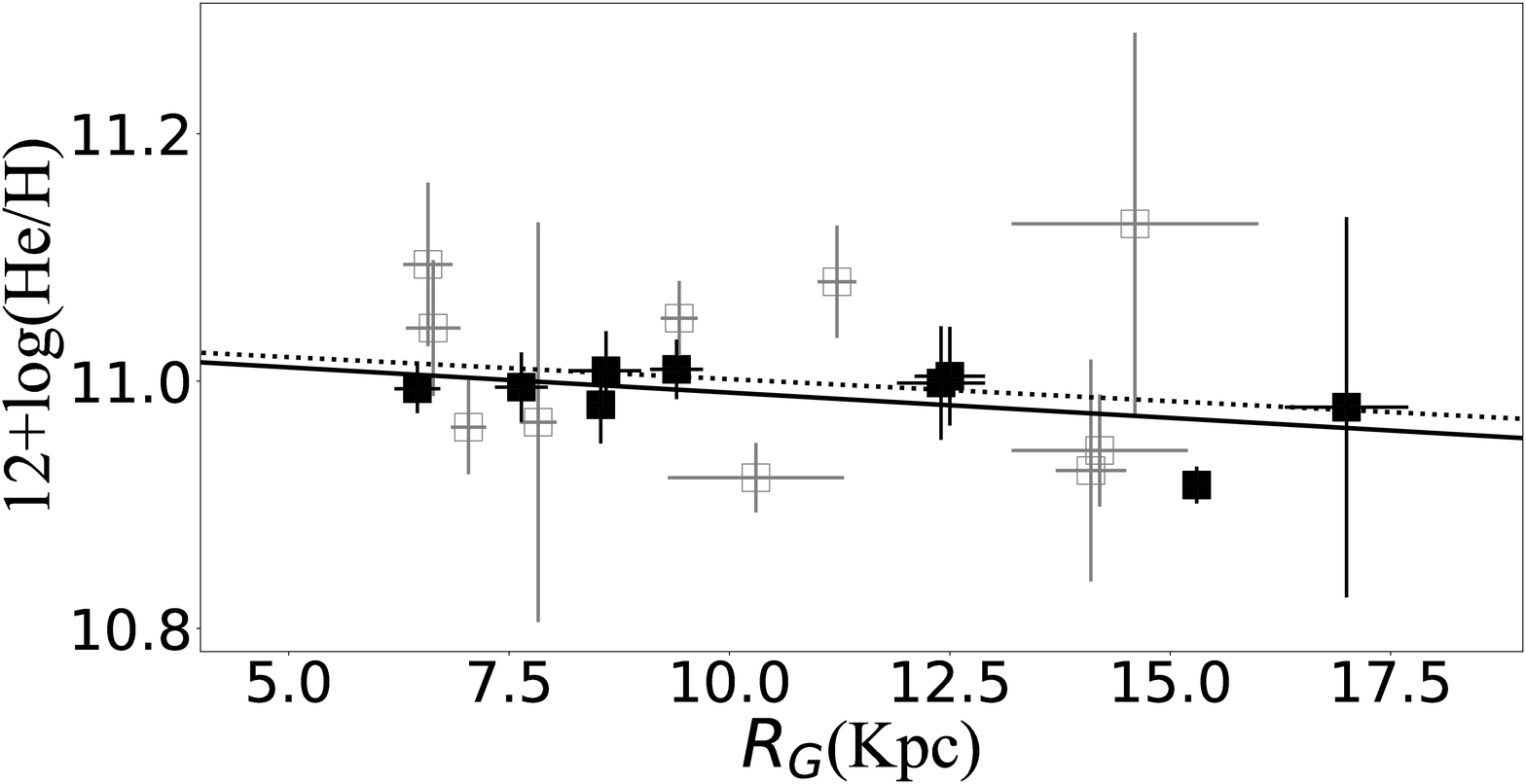} 
\includegraphics[scale=0.17]{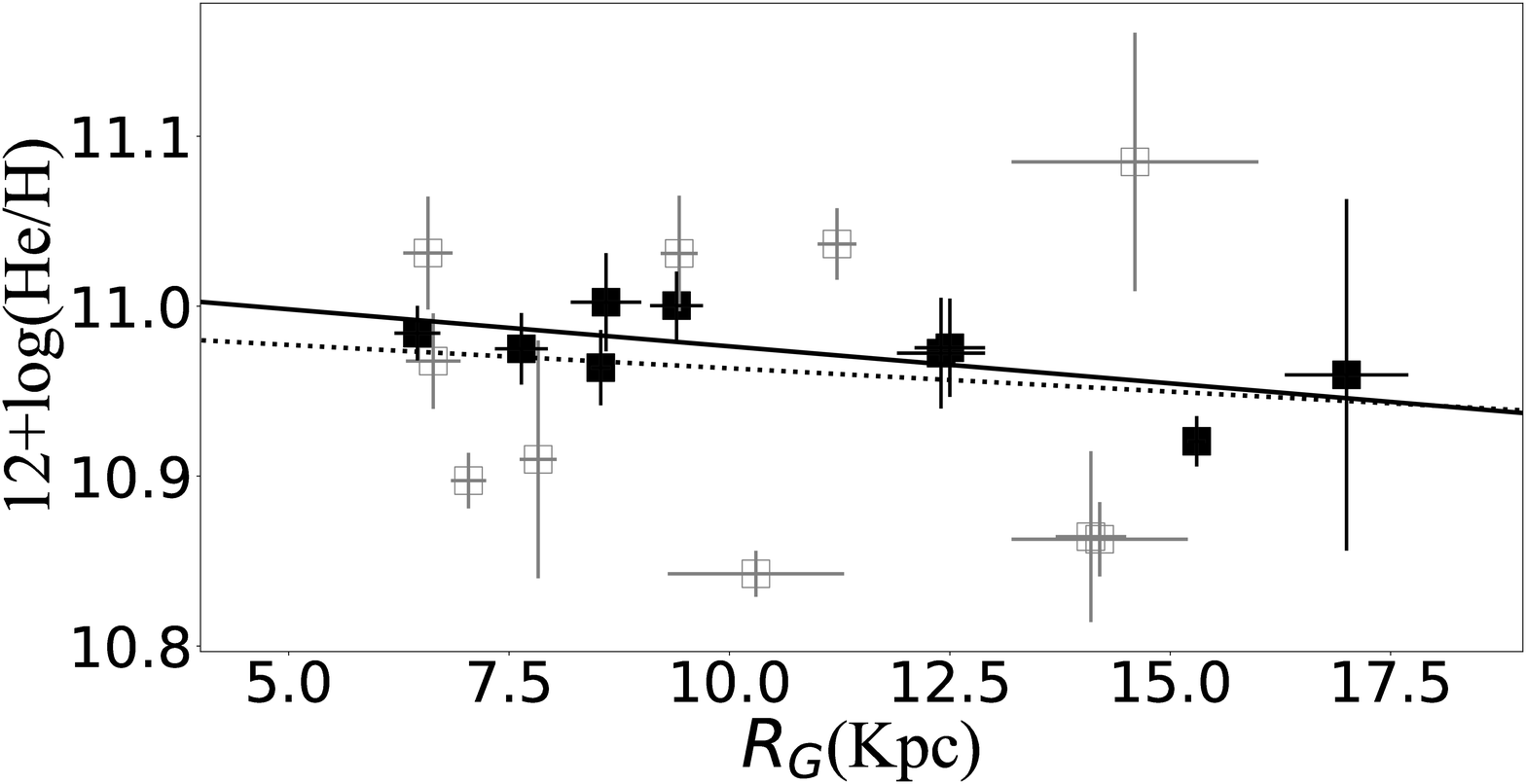} 
\\
\includegraphics[scale=0.17]{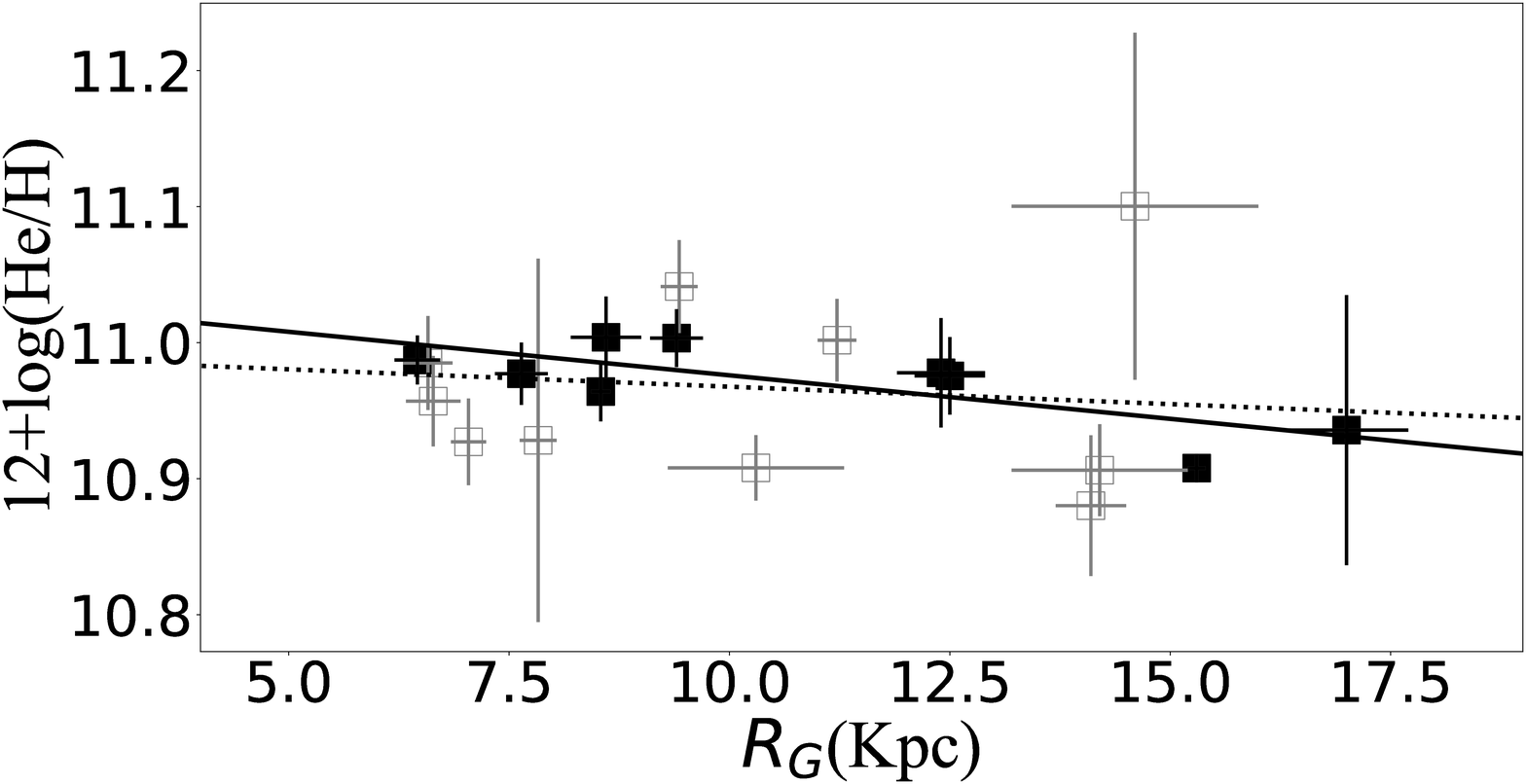} 
\includegraphics[scale=0.17]{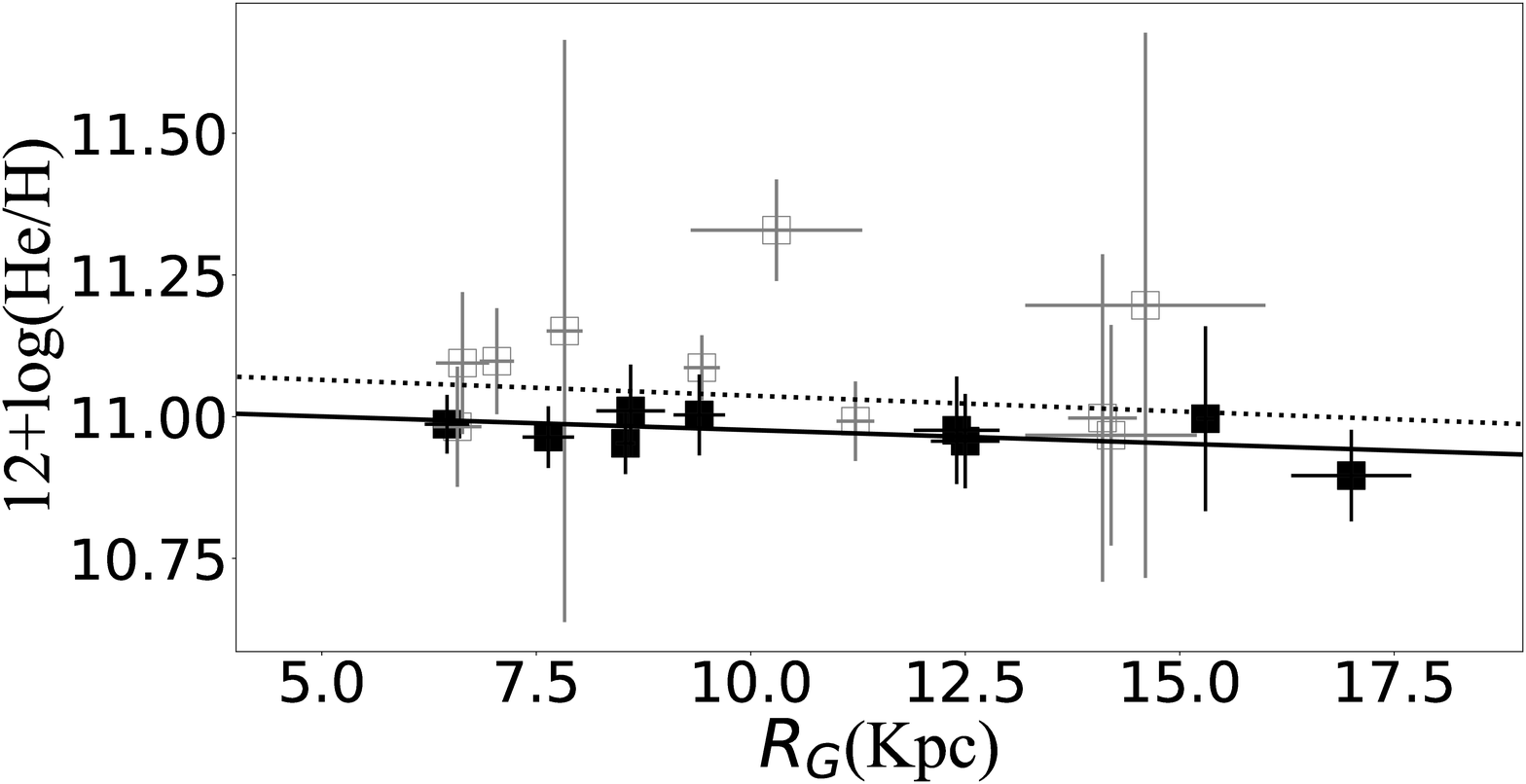} 
 \caption{Radial distribution of the He abundances of the {\hii} regions of our sample using different ICF(He) schemes. Upper left: using the ICF(He) proposed by \citet[][PTP77]{peimberttorrespeimbert77}; upper right: ICF(He) by \citet[][ZL03]{zhangliu03}; lower left: ICF(He) by \citet[][KS83]{kunthsargent83}; lower right: ICF(He) by \citet[][B09]{bresolinetal09}. Black full squares indicate {\hii} regions with log($\eta$) $<$ 0.9 and grey empty squares indicate {\hii} regions with log($\eta$)  $\geq$ 0.9. Solid lines represent the least-quares linear fits only to the objects with log($\eta$) $<$ 0.9 and dot lines represent the linear least-quares fits to all the objects included in each panel.}
 \label{fig:gradHe_icfs} 
 \end{figure*} 

\begin{figure} 
\centering 
\includegraphics[scale=0.18]{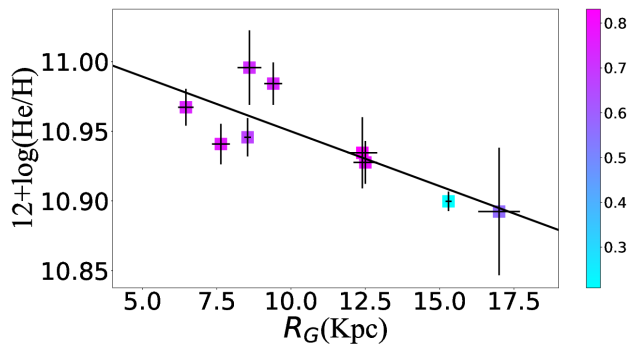} 
 \caption{Radial distribution of the He abundances of the {\hii} regions with log($\eta$) $<$ 0.9 assuming no ICF(He) (i.e. He/H =  \ionic{He}{+}/\ionic{H}{+}). The colour of the squares indicates the value of  log($\eta$) of each object. The solid lines represent the least-quares linear fits to the data.} 
 \label{fig:gradHe_noicf} 
 \end{figure} 

\begin{figure} 
\centering 
\includegraphics[scale=0.17]{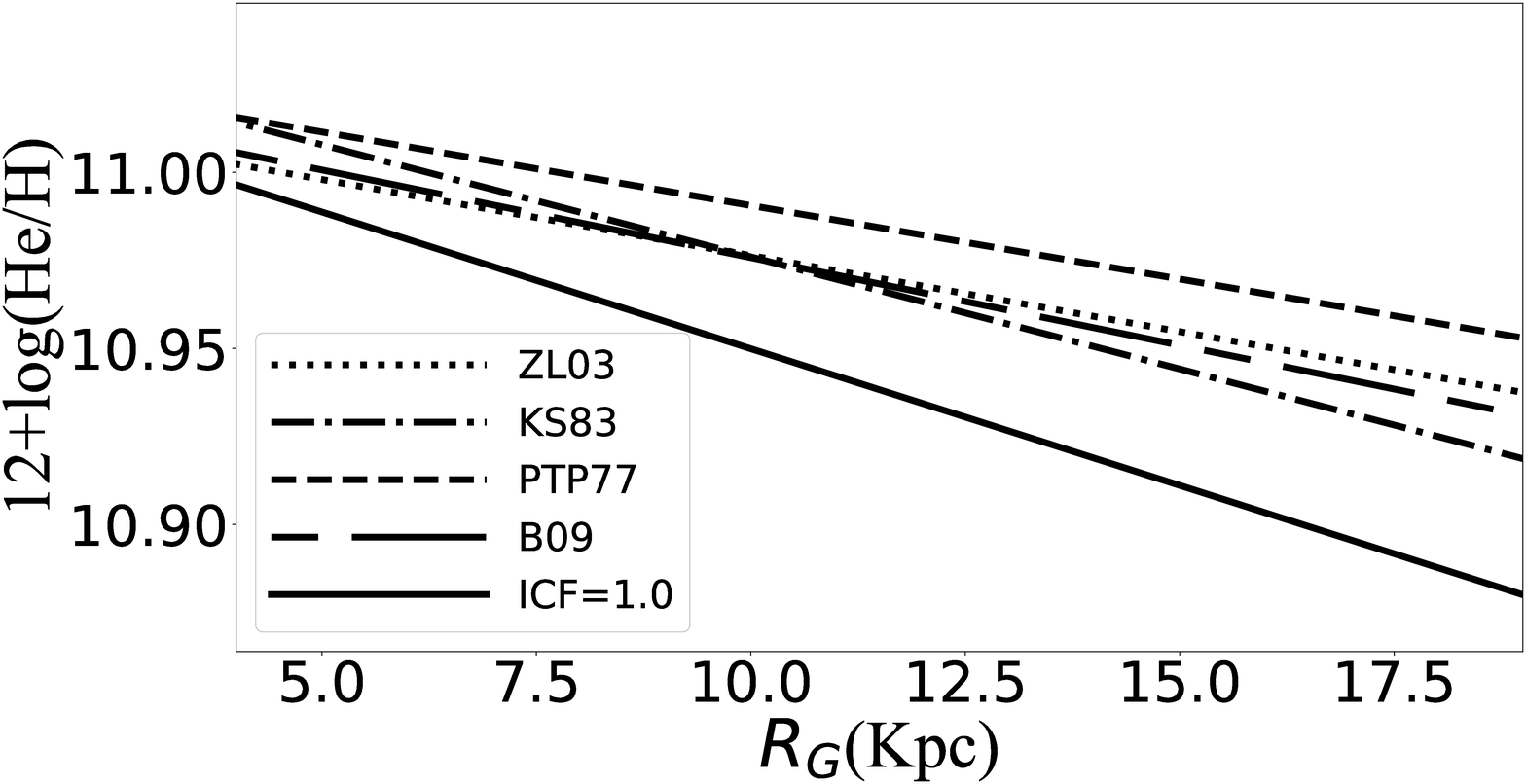} 
 \caption{Comparison of the least-quares linear fits of the He abundances obtained for the {\hii} regions with log($\eta$) $<$ 0.9 assuming different ICF(He) schemes or no ICF(He). Solid line: no ICF(He); dash line: ICF(He) by \citet[][PTP77]{peimberttorrespeimbert77}; dot line: ICF(He) by \citet[][ZL03]{zhangliu03}; dash dot line: ICF(He) by \citet[][KS83]{kunthsargent83}; long-dash dash line: ICF(He) by \citet[][B09]{bresolinetal09}.} 
 \label{fig:gradHe_comp} 
 \end{figure} 
 
\section{The Galactic radial abundance gradient of Helium} 
\label{sec:gradHe}

In Fig.~\ref{fig:gradHe_icfs} we show the radial distribution of the He abundances of the {\hii} regions of our sample -- a total of 19 objects -- using the different ICF(He) schemes introduced in equations \ref{eq:PTP77} to \ref{eq:B09}. We have also included the ring nebulae around O-type objects in this group because they do not contain chemically enriched material from the central star \citep[see][]{estebanetal16}. In these diagrams we have separated the nebulae in two groups, those with log($\eta$) $<$ 0.9 -- which are expected to contain a small fraction of \ionic{He}{0}  -- represented with full black squares and the rest of them (grey empty squares). This separation is made in order to explore the radial distribution of the He abundances minimizing the effect of the ICF(He) and maintaining a reasonable number of objects distributed along a significant part of the Galactic disc. 
The least-quares linear fits to the objects with log($\eta$) $<$ 0.9 (solid lines) and to all the objects included in each panel (dotted lines) are also included in each panel of Fig.~\ref{fig:gradHe_icfs}. The parameters of the fits along with their associated uncertainties are
shown in Table~\ref{tab:gradHe}. The uncertainties of the slope and intercept of the linear fits are computed through Monte Carlo simulations. We generate 10,000 random values of $R_{\rm G}$ and the He/H ratios for each observational point assuming a Gaussian distribution with a sigma equal to the uncertainty of each quantity. We performed a least-squares linear fit to each of these  random distributions. It is important to remark that the uncertainty of $R_{\rm G}$ has been considered in the fittings, which is not usually taken into account in most works except, for example, in \citet{estebanetal17} or \citet{estebangarciarojas18}. The uncertainties associated to the slope and intercept correspond to the standard deviation of the values of these two quantities obtained from the fits. From the radial gradients represented in Fig.~\ref{fig:gradHe_icfs} and their parameters collected in Table~\ref{tab:gradHe}, we can see that although the uncertainties of the slopes are larger than the slope values in most cases, they show pretty similar values in all cases, especially when only the objects with log($\eta$) $<$ 0.9 are considered. More importantly, the gradients are always negative independently on the ICF(He) scheme used. 

 In Fig.~\ref{fig:gradHe_noicf} we show the radial distribution and the least-quares linear fit of the He abundances of the {\hii} regions with log($\eta$) $<$ 0.9 assuming no ICF(He), i.e. He/H =  \ionic{He}{+}/\ionic{H}{+}. Each data point is represented with a colour proportional to its log($\eta$), and the diagram indicates that the value of the $\eta$ parameter does not seem to determine the behaviour of the gradient. In fact, the objects with the lowest values of log($\eta$) (i.e. those with the lowest expected fraction of \ionic{He}{0}) are precisely those located at larger $R_{\rm G}$ showing lower \ionic{He}{+}/\ionic{H}{+} ratios. In Fig.~\ref{fig:gradHe_comp}, we can compare the least-squares linear fits of the He abundances obtained for the {\hii} regions with log($\eta$) $<$ 0.9 assuming different ICF(He) schemes or no ICF(He). In the figure we can see that the slopes are rather similar in all cases, going from $-$0.0078 to $-$0.0044 dex kpc$^{-1}$ , with a dispersion of values smaller than the typical uncertainty of each individual determination of the slope. The differences are larger when we compare the values of the He/H given by each fit for a given $R_{\rm G}$. The no-ICF(He) case gives the lowest values of the He/H ratio  -- as expected -- and the ICF(He) by \citet{peimberttorrespeimbert77} gives the largest ones. It is interesting to note that the other ICF(He) schemes give He/H values that differ less than 0.02 dex. It is remarkable that the ICF scheme from \citet{bresolinetal09} is the most sensitive to the propagation of the errors. Its use requires very precise determinations of \ionic{O}{+}, \ionic{S}{+}, \ionic{O}{2+} and \ionic{S}{2+}.

The presence of abundance gradients of heavy elements -- such as O, N, Ne, S, Ar or Cl -- along the disc of the galaxy is a well established fact from the observational point of view \citep[e.g.][]{shaveretal83, deharvengetal00, rudolphetal06, balseretal11, estebangarciarojas18}.  However, although theoretical models predict its existence, this has not 
been the case for He. Very recent cosmological chemodynamical simulations for galaxy formation and evolution predict negative radial gradients of He/H due to the inside-out growth of galaxy discs as a function of time \citep{vincenzoetal19}. Previous chemical evolution models of the Milky Way disc by \citet{matteuccichiappini99} and 
\citet{chiappinietal02} predicted He/H gradients with slopes between $-$0.0085 and $-$0.002 dex kpc$^{-1}$ in the $R_{\rm G}$ range from 4 to 18 kpc. \citet{kubryketal15} obtain a similar range of values (from about $-$0.007 to $-$0.003 dex kpc$^{-1}$) from models including radial motions of gas and stars in the Milky Way disc and different sets of stellar yields. These values are pretty consistent with our results given in Table~\ref{tab:gradHe}. From the observational point of view, the evidence of a Galactic radial gradient of the He/H has been elusive. \citet{peimbertetal78} found a negative gradient of $-$0.02 $\pm$ 0.01 dex kpc$^{-1}$ from the analysis of the abundance obtained from 3 {\hei} lines in a small sample of 5 Galactic {\hii} regions. \citet{talentdufour79} obtain a marginal evidence of a negative gradient, $-$0.008 $\pm$ 0.008 dex kpc$^{-1}$ for a larger sample of objects based solely on the intensity of {\hei} {\lamb}5876 and \lamb6678 lines. Other works as \citet{hawley78}, \citet{shaveretal83}, \citet{deharvengetal00} or \citet{fernandezmartinetal17} find a flat He abundance distribution along the Galactic disc from observations of a limited number of {\hei} lines in samples with different number of objects. In their work,  \citet{fernandezmartinetal17} include some of the {\hii} regions of our sample (Sh2-83, Sh2-212 and NGC 7635), whose data are based on other spectroscopic  observations. The \ionic{He}{+}/\ionic{H}{+} ratios per line shown in their Table A.2. are consistent with our results, although there are slight differences attributable to the different recombination coefficients used. However, the methodology followed by \citet{fernandezmartinetal17} to adopt a final \ionic{He}{+}/\ionic{H}{+} value differs from ours, among other things, by the use of fewer {\hei } lines and the inclusion of {\hei } {\lamb}7065. Although their sample includes spectra of some Galactic {\hii} regions not considered in the present paper, we decided not to include them for the sake of having a group of nebulae with a homogeneous data reduction process; furthermore, the $R_{\rm G}$ of those objects are well covered by the {\hii} regions of our sample.

Other determinations of the radial He/H gradient in the Milky Way come from the analysis of the emission-line spectra of Planetary Nebulae (PNe). Early works such as those by \citet{dodoricoetal76}, \citet{peimbertserrano80} or \citet{faundez-abansmaciel87} found slopes between $-$0.03 and $-$0.02 dex kpc$^{-1}$, while others reported negligible gradients \citep[e.g.][]{pasqualiperinotto93}. At any rate, PNe are not confident probes of the abundances of the ISM  because they are composed of stellar ejecta material and can be contaminated by the products of nucleosynthesis. Trying to alleviate this drawback, \citet{maciel01} introduced corrections to the measured He abundance owing to the contamination from the progenitor star obtaining an essentially flat radial He/H gradient for a large sample of PNe. 

\section{Abundance deviations in WR ring nebulae} 
\label{sec:HeWR}

\begin{table*}
\centering
\caption{He, O and N abundance deviations with respect to abundance gradients in WR ring nebulae.} 
\label{tab:WRN} 
\begin{tabular}{lcccccc}
\hline
Object & 12+log(He/H) & $\Delta$(He/H) & 12+log(O/H) & $\Delta$(O/H) & 12+log(N/H) & $\Delta$(N/H) \\
\hline
G2.4+1.4  & 11.04 $\pm$ 0.02 & $+$0.04 $\pm$ 0.09 & 8.48 $\pm$ 0.25 & $-$0.09 $\pm$ 0.34 & 7.98 $\pm$ 0.23 & $+$0.10 $\pm$ 0.27 \\
NGC 6888 & 11.24 $\pm$ 0.01 & $+$0.25 $\pm$ 0.11 & 8.14 $\pm$ 0.19 & $-$0.33 $\pm$ 0.24 & 8.41 $\pm$ 0.30 & $+$0.67 $\pm$ 0.36 \\
RCW 58 & 11.23 $\pm$ 0.02 & $+$0.24 $\pm$ 0.10 & 8.60 $\pm$ 0.16 & $+$0.11 $\pm$ 0.20 & 8.67 $\pm$ 0.18 & $+$0.91 $\pm$ 0.23 \\
Sh~2-298 &  11.08 $\pm$ 0.09 & $+$0.12 $\pm$ 0.15 & 8.43 $\pm$ 0.06 & $+$0.10 $\pm$ 0.15 & 7.60 $\pm$ 0.13 & $+$0.07 $\pm$ 0.20 \\
Sh~2-308 & 11.22 $\pm$ 0.01 & $+$0.24 $\pm$ 0.12 & 7.89 $\pm$ 0.23 & $-$0.51 $\pm$ 0.28 & 8.05: & $+$0.41:  \\
\hline
\end{tabular}
\end{table*} 

Ring nebulae are bubbles of gas swept-up by the mechanical action of the mass loss episodes experienced by their massive stellar progenitors. They are 
rather common around Wolf-Rayet (WR) stars and luminous blue variables (LBVs). There have been several spectroscopical works devoted to study the chemical composition of the ionized gas contained in ring nebulae around Galactic WR stars \citep[e.g.][]{kwitter84, estebanetal92, estebanetal16, stocketal11}. Those works have found that some ring nebulae (NGC~6888, RCW~58,  Sh~2-308 and M~1-67) show clear chemical enrichment patterns, in general they are He and N enriched and 
some of them show some O deficiency, indicating that they are composed by ejecta material that has suffered contamination by the CNO cycle. \citet{estebanvilchez92} and \citet{estebanetal92} compared the abundance pattern of Galactic ejecta ring nebulae with the surface abundances predicted by  evolutionary models of massive stars, finding that the He, N and O abundances of the nebulae are consistent with the expected surface composition of stars with initial masses between 25 and 40 M$_\odot$ at the red supergiant (RSG) phase, prior to the onset of the WR stage. Later studies by \citet{mesadelgadoetal14} and \citet{estebanetal16} using stellar evolution models by \citet{ekstrometal12} and \citet{georgyetal12} confirmed that non-rotational models of stars of initial masses between 25 and 40 M$_\odot$ seem to reproduce the abundance patters in most of the ejecta nebulae, being this range wider in the case of Sh~2-308, from 25 to 50 $M_{\odot}$. Only rotational models of  25 $M_{\odot}$ show agreement with the data for NGC~6888, RCW~58 and Sh~2-308.

The availability of new Gaia distances, the results of this paper and the recent reassessment of the Galactic radial gradients of O/H and N/H by \citet{estebangarciarojas18}, allow us to perform a better estimate of the chemical enrichment pattern in He, N and O of the WR ring nebulae observed by our group \citep{estebanetal16}. In Table~\ref{tab:WRN} we give the total He/H, O/H and N/H ratios and the difference with the values expected from the radial abundance gradients for the sample of WR ring nebulae included in Table~\ref{tab:sample}. The He abundances of each object have been taken from Table~\ref{tab:he}. In the case of RCW~58, 
Sh~2-298 and Sh~2-308, we have considered the mean and standard deviation of the 4 values obtained using the different ICF(He) schemes. The O abundances have been calculated simply adding the \ionic{O}{+}/\ionic{H}{+} and \ionic{O}{2+}/\ionic{H}{+} ratios given in Table~\ref{tab:ne_te_abund}, except in the case of G2.4$+$1.4, the only object where we find a substantial amount of  \ionic{He}{2+}/\ionic{H}{+}. In this case, we have considered the ICF(O) proposed by \citet{delgadoingladaetal14} for determining the O/H ratio. The N/H ratio has been calculated using the  \ionic{N}{+} abundance given in Table~\ref{tab:ne_te_abund} and  the classical ICF(N) by \citet{peimbertcostero69}, that assumes N/O =  \ionic{N}{+}/ \ionic{O}{+}. 

In Table~\ref{tab:WRN} we can see that G2.4$+$1.4 and Sh~2-298 do not show He/H, O/H and N/H ratios significantly larger than the expected values from the radial gradients considering the abundance uncertainties. NGC~6888, RCW~58 and Sh~2-308 show clear overabundances of He/H and N/H, although the errors in the last nebula are rather large due to its highly uncertain \ionic{O}{+}/\ionic{H}{+} ratio. The overabundance of He/H is very similar in the three objects, of about +0.24 $\pm$ 0.11 dex. Only NGC~6888 and Sh~2-308 show some possible oxygen deficiency, a result that would indicate that 
ON-cycle of the CNO burning has been more effective than CN-cycle in the interior of the massive progenitors stars of the nebulae \citep{estebanetal16}. These results confirm that only  NGC~6888, RCW~58 and Sh~2-308 should be considered ejecta nebulae.

\section{Conclusions}\label{conclusions}     
We determine the radial abundance gradient of helium of the Milky Way from published spectra of {\hii} regions. The data set is the largest collection of deep spectra of Galactic {\hii} regions available. We also include data of similar quality for several ring nebulae surrounding massive O and WR stars. The total number of nebulae included in the sample is 24. We have revised the Galactocentric distances of the objects, $R_{\rm G}$, considering {\it Gaia} DR2 parallaxes and previous kinematic and spectroscopical determinations.  We have determined the physical conditions --{\elecd} and {\elect} -- and the ionic abundance of \ionic{He}{+} and  other selected ions in a homogeneous way and using the most recent atomic data sets. We have determined the \ionic{He}{+} abundance using several {\hei} recombination lines. In the case of the {\hii} regions, we have used between 3 and 10 individual {\hei} lines depending on the object, selecting only those well-measured lines that are not affected by line-blending, telluric contamination or important self-absorption effects. The total He abundance of the objects have been estimated using four different ICF(He) schemes based on different ionic ratios. We have determined the Galactic radial He abundance gradient using the results of 19 objects, including {\hii} regions and ring nebulae around O-type stars. We find that although the uncertainties of the slopes are larger than the slope values obtained using several of the four ICF(He) schemes, they show consistent values in all cases, especially when only the objects with log($\eta$) $<$ 0.9 are considered. More importantly, the gradients are always negative independently on the ICF(He) scheme used. The slope values go from $-$0.0078 to $-$0.0044 dex kpc$^{-1}$, consistent with the predictions of different chemical evolution models of  the Milky Way and chemodynamical simulations of galactic discs. We have estimated the abundance deviations of He, O and N of the ring nebulae around WR stars included in our sample with respect to the radial gradients, finding that only NGC~6888, RCW~585 and Sh~2-308 can be considered ejecta nebulae. These objects show He and N overabundances. The degree of enrichment of He is very similar and about +0.24 $\pm$ 0.11 dex in the three objects. NGC~6888 and Sh~2-308 show a possible O deficiency. 

\section*{Acknowledgements} We thank the referee \'Angeles I. D\'iaz for a constructive report. We acknowledge support from the State Research Agency (AEI) of the Spanish Ministry of Science, Innovation and Universities (MCIU) and the European Regional Development Fund (FEDER) under grant with reference AYA2015-65205-P. JG-R acknowledges support from an Advanced Fellowship from the Severo Ochoa excellence program (SEV-2015-0548). The authors acknowledge support under grant P/308614 financed by funds transferred from the Spanish Ministry of Science, Innovation and Universities, charged to the General State Budgets and with funds transferred from the General Budgets of the Autonomous Community of the Canary Islands by the MCIU. KZA-C acknowledges support from Mexican CONACYT posdoctoral grant 364239. JEM-D thanks the Fundaci\'on Carolina for the support provided for his Master's studies. JEM-D also thanks the support of the Instituto de Astrof\'isica de Canarias under the Astrophysicist Resident Program and acknowledges support from the Mexican CONACyT (grant CVU 602402).
 This work has made use of data from the European Space Agency (ESA) mission
{\it Gaia} (\url{https://www.cosmos.esa.int/gaia}), processed by the {\it Gaia}
Data Processing and Analysis Consortium (DPAC,
\url{https://www.cosmos.esa.int/web/gaia/dpac/consortium}). Funding for the DPAC
has been provided by national institutions, in particular the institutions
participating in the {\it Gaia} Multilateral Agreement.

\section*{Data availability}

This research is based on public data available in the references. The whole dataset can be obtained from the authors by request. All the results are available in the tables and in the appendix of this article.




\bibliographystyle{mnras} 
\bibliography{cesar_bibliography} 


\appendix

\section{Ionic abundances for each individual {\hei} line}
\label{appex:1}
In this Appendix we include four tables. Table~A1 shows the \ionic{He}{+}/\ionic{H}{+} ratio for each individual {\hei} line of the spectra analyzed in this work. In the second column we indicate the level state that produce each line, S for singlet and T for triplet. We also include the mean \ionic{He}{+}/\ionic{H}{+} ratio for each spectra. In Table~A2  we give the list of individual {\hei} lines used to derive the mean value of the \ionic{He}{+}/\ionic{H}{+} ratio adopted for each spectrum. In table  Table~A3 we include the results of the abundances of \ionic{He}{+}/\ionic{H}{+} estimated with the code \texttt{Helio14}. In Table~A4 we include the ICFs used in each total abundance calculation. 

 \begin{table*}
 \centering
 \caption{ \ionic{He}{+}/\ionic{H}{+} ratios for each individual {\hei} line of all the spectra used in this work}
\label{tab:hep}
\begin{tabular}{ccccccc}
\hline       
{\hei} line & Level &  \multicolumn{5}{c}{G2.4+1.4} \\                                       
(\AA) & state & A1 & A2 & A3 & A4 & A5\\  
\hline  
$\lambda$4471& T & - & 10.865 $\pm$ 0.109& - & - & - \\
$\lambda$5876& T & 10.808 $\pm$ 0.032& 10.922 $\pm$ 0.029& 10.817 $\pm$ 0.089& 10.925 $\pm$ 0.025& 11.003 $\pm$ 0.057\\
$\lambda$6678& S & 10.720 $\pm$ 0.118& 10.896 $\pm$ 0.036& - & 10.867 $\pm$ 0.039& 10.838 $\pm$ 0.093\\
\multicolumn{2}{c}{Mean} & 10.800 $\pm$ 0.065& 10.909 $\pm$ 0.016& 10.817 $\pm$ 0.089& 10.906 $\pm$ 0.031& 10.937 $\pm$ 0.071\\     
\hline                                     
 &  & M8& M16& M17& M20& M42 \\  
\hline  
$\lambda$3614& S & 10.844 $\pm$ 0.052& 10.919 $\pm$ 0.052& 10.934 $\pm$ 0.108& 10.868 $\pm$ 0.057& 10.947 $\pm$ 0.031\\
$\lambda$3965& S & 10.849 $\pm$ 0.035& 10.883 $\pm$ 0.017& 10.938 $\pm$ 0.022& 10.860 $\pm$ 0.017& 10.925 $\pm$ 0.013\\
$\lambda$4026& T & - & - & - & - & - \\
$\lambda$4121& T & 10.859 $\pm$ 0.027& 11.013 $\pm$ 0.044& 11.068 $\pm$ 0.056& 10.598 $\pm$ 0.110& 11.051 $\pm$ 0.019\\
$\lambda$4388& S & 10.830 $\pm$ 0.013& 10.886 $\pm$ 0.017& 10.961 $\pm$ 0.026& 10.826 $\pm$ 0.026& 10.929 $\pm$ 0.009\\
$\lambda$4438& S & 10.880 $\pm$ 0.044& 10.817 $\pm$ 0.103& 10.979 $\pm$ 0.109& 10.938 $\pm$ 0.109& 10.946 $\pm$ 0.035\\
$\lambda$4471& T & 10.840 $\pm$ 0.013& 10.893 $\pm$ 0.013& 10.973 $\pm$ 0.013& 10.850 $\pm$ 0.013& 10.950 $\pm$ 0.005\\
$\lambda$4713& T & 10.839 $\pm$ 0.014& 10.900 $\pm$ 0.019& 11.013 $\pm$ 0.026& 10.825 $\pm$ 0.027& 11.109 $\pm$ 0.011\\
$\lambda$4922& S & 10.829 $\pm$ 0.013& 10.883 $\pm$ 0.013& 10.959 $\pm$ 0.013& 10.862 $\pm$ 0.013& 10.946 $\pm$ 0.005\\
$\lambda$5016& S & 10.810 $\pm$ 0.013& 10.864 $\pm$ 0.013& 10.926 $\pm$ 0.013& 10.849 $\pm$ 0.013& 10.900 $\pm$ 0.006\\
$\lambda$5048& S & 11.001 $\pm$ 0.018& 11.057 $\pm$ 0.053& 10.807 $\pm$ 0.048& 10.925 $\pm$ 0.048& 11.463 $\pm$ 0.011\\
$\lambda$5876& T & 10.838 $\pm$ 0.013& 10.903 $\pm$ 0.013& 10.974 $\pm$ 0.013& 10.855 $\pm$ 0.014& 10.990 $\pm$ 0.014\\
$\lambda$6678& S & 10.845 $\pm$ 0.013& 10.895 $\pm$ 0.018& 10.988 $\pm$ 0.017& 10.862 $\pm$ 0.018& 10.965 $\pm$ 0.026\\
$\lambda$7281& S & 10.808 $\pm$ 0.016& 10.900 $\pm$ 0.023& 10.928 $\pm$ 0.022& 10.885 $\pm$ 0.022& 10.872 $\pm$ 0.038\\
$\lambda$9464& T & 10.751 $\pm$ 0.022& 10.690 $\pm$ 0.039& - & - & 10.889 $\pm$ 0.065\\
\multicolumn{2}{c}{Mean} & 10.838 $\pm$ 0.009& 10.892 $\pm$ 0.009& 10.967 $\pm$ 0.013& 10.854 $\pm$ 0.009& 10.946 $\pm$ 0.014\\
\hline
& & & & & \multicolumn{2}{c}{NGC 6888} \\
& & NGC 2579& NGC 3576& NGC 3603& A1& A2 \\
\hline
$\lambda$3614& S & 10.730 $\pm$ 0.052& 10.927 $\pm$ 0.039& - & - & - \\
$\lambda$3965& S & 10.904 $\pm$ 0.026& 10.925 $\pm$ 0.013& 10.895 $\pm$ 0.048& - & - \\
$\lambda$4026& T & 10.962 $\pm$ 0.022& 10.951 $\pm$ 0.013& - & - & 11.191 $\pm$ 0.054\\
$\lambda$4121& T & 11.050 $\pm$ 0.048& - & 11.202 $\pm$ 0.086& - & 11.600 $\pm$ 0.134\\
$\lambda$4388& S & 10.961 $\pm$ 0.026& 10.944 $\pm$ 0.013& 11.025 $\pm$ 0.044& - & 11.087 $\pm$ 0.068\\
$\lambda$4438& S & 10.965 $\pm$ 0.070& 11.003 $\pm$ 0.044& - & - & - \\
$\lambda$4471& T & 10.964 $\pm$ 0.017& 11.040 $\pm$ 0.009& 11.021 $\pm$ 0.013& - & 11.079 $\pm$ 0.014\\
$\lambda$4713& T & 11.062 $\pm$ 0.023& 11.086 $\pm$ 0.010& 11.106 $\pm$ 0.032& - & - \\
$\lambda$4922& S & 10.915 $\pm$ 0.017& 10.938 $\pm$ 0.009& 10.971 $\pm$ 0.018& - & 11.143 $\pm$ 0.026\\
$\lambda$5016& S & 10.898 $\pm$ 0.017& 10.900 $\pm$ 0.009& 10.876 $\pm$ 0.022& 10.861 $\pm$ 0.122& 11.080 $\pm$ 0.022\\
$\lambda$5048& S & - & - & - & - & 11.176 $\pm$ 0.084\\
$\lambda$5876& T & 11.011 $\pm$ 0.018& 10.900 $\pm$ 0.018& 10.999 $\pm$ 0.018& 11.049 $\pm$ 0.036& 11.161 $\pm$ 0.014\\
$\lambda$6678& S & 10.943 $\pm$ 0.017& 10.946 $\pm$ 0.026& 10.974 $\pm$ 0.022& 10.865 $\pm$ 0.073& 11.151 $\pm$ 0.018\\
$\lambda$7281& S & - & 10.953 $\pm$ 0.031& 10.972 $\pm$ 0.029& - & 10.986 $\pm$ 0.071\\
$\lambda$9464& T & 10.949 $\pm$ 0.018& 11.031 $\pm$ 0.048& 11.226 $\pm$ 0.040& - & - \\
\multicolumn{2}{c}{Mean} & 10.934 $\pm$ 0.026& 10.941 $\pm$ 0.015& 10.996 $\pm$ 0.027& 10.977 $\pm$ 0.088& 11.154 $\pm$ 0.014\\
\hline
& & \multicolumn{4}{c}{NGC 6888} & NGC 7635\\
& &  A3& A4 & A5 & A6 & A1\\
\hline
$\lambda$4026& T & 11.287 $\pm$ 0.070& 11.240 $\pm$ 0.090& 11.197 $\pm$ 0.065& 11.206 $\pm$ 0.046& 10.818 $\pm$ 0.137\\
$\lambda$4388& S & 11.184 $\pm$ 0.131& - & 11.128 $\pm$ 0.098& 11.130 $\pm$ 0.107& 10.879 $\pm$ 0.160\\
$\lambda$4471& T & 11.156 $\pm$ 0.016& 11.249 $\pm$ 0.045& 11.235 $\pm$ 0.019& 11.250 $\pm$ 0.020& 10.818 $\pm$ 0.015\\
$\lambda$4713& T & 11.068 $\pm$ 0.043& 11.108 $\pm$ 0.049& 11.138 $\pm$ 0.068& 11.121 $\pm$ 0.043& 10.872 $\pm$ 0.064\\
$\lambda$4922& S & 11.215 $\pm$ 0.018& 11.268 $\pm$ 0.030& 11.237 $\pm$ 0.062& 11.243 $\pm$ 0.034& 10.833 $\pm$ 0.028\\
$\lambda$5016& S & 11.174 $\pm$ 0.017& 11.135 $\pm$ 0.022& 11.194 $\pm$ 0.022& 11.191 $\pm$ 0.022& 10.881 $\pm$ 0.015\\
$\lambda$5048& S & 11.122 $\pm$ 0.087& 11.071 $\pm$ 0.121& 11.103 $\pm$ 0.125& 11.100 $\pm$ 0.068& 10.953 $\pm$ 0.108\\
$\lambda$5876& T & 11.237 $\pm$ 0.016& 11.255 $\pm$ 0.013& 11.235 $\pm$ 0.034& 11.271 $\pm$ 0.012& 10.889 $\pm$ 0.010\\
$\lambda$6678& S & 11.227 $\pm$ 0.023& 11.237 $\pm$ 0.019& 11.257 $\pm$ 0.022& 11.248 $\pm$ 0.028& 10.878 $\pm$ 0.012\\
\multicolumn{2}{c}{Mean} & 11.228 $\pm$ 0.013& 11.251 $\pm$ 0.010& 11.240 $\pm$ 0.013& 11.259 $\pm$ 0.016& 10.879 $\pm$ 0.005\\
\end{tabular}
\end{table*}


\setcounter{table}{0}
 \begin{table*}
\centering
\caption{continued}
\begin{tabular}{ccccccc}
\hline       
{\hei} line & Level &  \multicolumn{5}{c}{NGC 7635} \\                                       
(\AA) & state & A2 & A3 & A4 & A5 & A6\\  
\hline  
$\lambda$4026& T & - & 10.780 $\pm$ 0.137& 10.811 $\pm$ 0.028& 11.140 $\pm$ 0.131& - \\
$\lambda$4388& S &  - & - & 10.805 $\pm$ 0.039& - & - \\
$\lambda$4471& T & 10.899 $\pm$ 0.026& 10.914 $\pm$ 0.015& 10.795 $\pm$ 0.010& 10.887 $\pm$ 0.045& 10.823 $\pm$ 0.025\\
$\lambda$4713& T & 10.893 $\pm$ 0.083& 10.928 $\pm$ 0.057& 10.834 $\pm$ 0.022& - & - \\
$\lambda$4922& S & 10.992 $\pm$ 0.036& 10.997 $\pm$ 0.036& 10.804 $\pm$ 0.011& 10.884 $\pm$ 0.090& 10.744 $\pm$ 0.097\\
$\lambda$5016& S & - & - & 10.771 $\pm$ 0.029& 10.961 $\pm$ 0.025& 10.933 $\pm$ 0.031\\
$\lambda$5048& S & 11.034 $\pm$ 0.131& 10.918 $\pm$ 0.145& 10.887 $\pm$ 0.025& - & - \\
$\lambda$5876& T & 11.040 $\pm$ 0.012& 10.989 $\pm$ 0.014& 10.853 $\pm$ 0.009& 10.969 $\pm$ 0.019& 10.933 $\pm$ 0.016\\
$\lambda$6678& S & 11.066 $\pm$ 0.016& 10.993 $\pm$ 0.020& 10.849 $\pm$ 0.010& 10.946 $\pm$ 0.028& 10.896 $\pm$ 0.029\\
$\lambda$7281& S & - & 10.627 $\pm$ 0.139& 10.798 $\pm$ 0.046& 11.212 $\pm$ 0.119& - \\
\multicolumn{2}{c}{Mean} & 11.049 $\pm$ 0.013& 10.987 $\pm$ 0.013& 10.798 $\pm$ 0.009& 10.959 $\pm$ 0.015& 10.925 $\pm$ 0.015\\
\hline       
& &  RCW 52& RCW 58& Sh 2-83& Sh 2-100& Sh 2-127 \\
\hline
$\lambda$3965& S &  - & 11.484 $\pm$ 0.069& - & - & - \\
$\lambda$4026& T & - & 11.297 $\pm$ 0.068& - & 10.973 $\pm$ 0.033& - \\
$\lambda$4388& S &  - & 11.186 $\pm$ 0.108& - & 10.923 $\pm$ 0.038& - \\
$\lambda$4471& T & 10.761 $\pm$ 0.147& 11.226 $\pm$ 0.018& 10.907 $\pm$ 0.040& 10.986 $\pm$ 0.013& 10.746 $\pm$ 0.039\\
$\lambda$4713& T & - & 11.134 $\pm$ 0.144& - & - & - \\
$\lambda$4922& S &  - & 11.186 $\pm$ 0.066& 10.895 $\pm$ 0.037& 10.989 $\pm$ 0.024& 10.801 $\pm$ 0.026\\
$\lambda$5016& S &  10.827 $\pm$ 0.141& 11.220 $\pm$ 0.016& 10.883 $\pm$ 0.049& 10.919 $\pm$ 0.056& 10.808 $\pm$ 0.014\\
$\lambda$5048& S &  - & - & 10.999 $\pm$ 0.078& 10.980 $\pm$ 0.070& - \\
$\lambda$5876& T & 10.819 $\pm$ 0.024& 11.210 $\pm$ 0.016& 10.972 $\pm$ 0.014& 11.042 $\pm$ 0.024& 10.832 $\pm$ 0.012\\
$\lambda$6678& S &  10.840 $\pm$ 0.038& 11.257 $\pm$ 0.020& 10.902 $\pm$ 0.021& 11.008 $\pm$ 0.035& 10.790 $\pm$ 0.015\\
$\lambda$7281& S &  - & - & 10.778 $\pm$ 0.046& 10.897 $\pm$ 0.047& 10.561 $\pm$ 0.072\\
\multicolumn{2}{c}{Mean} & 10.825 $\pm$ 0.009& 11.217 $\pm$ 0.008& 10.899 $\pm$ 0.007& 10.984 $\pm$ 0.015& 10.799 $\pm$ 0.008\\
\hline       
& &  Sh 2-128& Sh 2-152& Sh 2-209& Sh 2-212& Sh 2-288 \\
\hline
$\lambda$4026& T & - & 10.817 $\pm$ 0.039& - & - & 10.853 $\pm$ 0.047\\
$\lambda$4388& S & - & 10.751 $\pm$ 0.097& - & - & - \\
$\lambda$4471& T & 10.924 $\pm$ 0.024& 10.839 $\pm$ 0.011& - & 10.975 $\pm$ 0.024& 10.780 $\pm$ 0.022\\
$\lambda$4713& T & - & 10.961 $\pm$ 0.052& - & - & - \\
$\lambda$4922& S & 10.950 $\pm$ 0.032& 10.807 $\pm$ 0.029& - & 11.036 $\pm$ 0.020& 10.739 $\pm$ 0.024\\
$\lambda$5016& S & 10.913 $\pm$ 0.013& 10.658 $\pm$ 0.019& 10.709 $\pm$ 0.079& 10.964 $\pm$ 0.039& 10.798 $\pm$ 0.016\\
$\lambda$5048& S & 11.113 $\pm$ 0.078& 10.792 $\pm$ 0.017& - & - & 10.745 $\pm$ 0.096\\
$\lambda$5876& T & 10.970 $\pm$ 0.011& 10.803 $\pm$ 0.009& 10.923 $\pm$ 0.027& 11.020 $\pm$ 0.014& 10.798 $\pm$ 0.026\\
$\lambda$6678& S & 10.945 $\pm$ 0.015& 10.790 $\pm$ 0.010& 10.882 $\pm$ 0.044& 10.765 $\pm$ 0.050& 10.745 $\pm$ 0.040\\
$\lambda$7281& S & 10.847 $\pm$ 0.031& 10.833 $\pm$ 0.066& 10.780 $\pm$ 0.074& - & 10.683 $\pm$ 0.083\\
\multicolumn{2}{c}{Mean} & 10.928 $\pm$ 0.015& 10.802 $\pm$ 0.007& 10.893 $\pm$ 0.046& 11.019 $\pm$ 0.019& 10.792 $\pm$ 0.010\\
\hline       
& &  Sh 2-298& Sh 2-308& Sh 2-311 & & \\
\hline
$\lambda$3614& S & - & - & 10.885 $\pm$ 0.065 & & \\
$\lambda$3965& S & 10.819 $\pm$ 0.102& - & 10.905 $\pm$ 0.018 & & \\
$\lambda$4026& T & 10.922 $\pm$ 0.048& - & 10.915 $\pm$ 0.017 & & \\
$\lambda$4121& T & - & - & 10.926 $\pm$ 0.039 & & \\
$\lambda$4388& S & 11.069 $\pm$ 0.091& - & 10.922 $\pm$ 0.017 & & \\
$\lambda$4438& S & - & - & 10.963 $\pm$ 0.065 & & \\
$\lambda$4471& T & 10.975 $\pm$ 0.016& - & 10.913 $\pm$ 0.013 & & \\
$\lambda$4713& T & 10.949 $\pm$ 0.079& - & 10.919 $\pm$ 0.018 & & \\
$\lambda$4922& S & 10.986 $\pm$ 0.034& - & 10.910 $\pm$ 0.017 & & \\
$\lambda$5016& S & 10.740 $\pm$ 0.074& - & 10.892 $\pm$ 0.013 & & \\
$\lambda$5048& S & - & - & 11.040 $\pm$ 0.044 & & \\
$\lambda$5876& T & 10.969 $\pm$ 0.017& 11.199 $\pm$ 0.047& 10.911 $\pm$ 0.018 & & \\
$\lambda$6678& S & 11.020 $\pm$ 0.037& 11.202 $\pm$ 0.077& 10.916 $\pm$ 0.017 & & \\
$\lambda$7281& S & 10.617 $\pm$ 0.101& - & 10.909 $\pm$ 0.022 & & \\
$\lambda$9464& T & - & - & 10.985 $\pm$ 0.035 & & \\
\multicolumn{2}{c}{Mean} & 10.971 $\pm$ 0.017& 11.200 $\pm$ 0.062& 10.914 $\pm$ 0.006\\
\hline
\end{tabular}
\end{table*}

 \begin{table*}
 \centering
 \caption{ List of lines used for the calculation of the mean \ionic{He}{+}/\ionic{H}{+} ratio for each spectra.}
\label{tab:list}
\begin{tabular}{lcc}
\hline       
Object& Zone & Lines used \\ 
\hline 
G2.4+1.4 & A1 &  5876, 6678 \\
& A2 &  4471, 5876, 6678 \\
& A3 &  5876 \\
& A4 & 5876, 6678 \\
& A5 &  5876, 6678 \\
M8 & & 3614, 3965, 4121, 4388, 4438, 4471, 4713, 4922, 5876, 6678 \\
M16 & & 3614, 3965, 4388, 4438, 4471, 4713, 4922, 5876, 6678, 7281 \\
M17 & & 3614, 3965, 4388, 4438, 4471, 4922, 5876, 6678 \\
M20 & & 3614, 3965, 4388, 4438, 4471, 4922, 5016, 5876, 6678 \\
M42 & & 3614, 3965, 4388, 4438, 4471, 4922, 5876, 6678, 7281, 9464 \\
NGC 2579  & & 3965, 4026, 4388, 4438, 4471, 4922, 5016, 6678, 9464 \\
NGC 3576 & & 3614, 3965, 4026, 4388, 4438, 4922, 6678, 7281, 9464 \\
NGC 3603 & & 4121, 4388, 4471, 4922, 5876, 6678, 7281 \\
NGC 6888 & A1 &  5016, 5876, 6678 \\
& A2 & 4026, 4388, 4922, 5048, 5876, 6678 \\
& A3 & 4026, 4388, 4922, 5876, 6678 \\
& A4 & 4026, 4471, 4922, 5876, 6678 \\
& A5 & 4026, 4471, 4922, 5876, 6678 \\
& A6 & 4026, 4471, 4922, 5876, 6678 \\
RCW 52 & & 4471, 5016, 5876 \\
RCW 58 & & 4388, 4471, 4922, 5016, 5876 \\
NGC 7635 & A1 &4026, 4388, 4713, 5016, 6678 \\
& A2 & 5048, 5876, 6678 \\
& A3 & 4713, 4922, 5048, 5876, 6678 \\
& A4 & 4026, 4388, 4471, 4922, 5016, 7281 \\
& A5 & 4922, 5016, 5876, 6678 \\
& A6 & 5016, 5876, 6678 \\
Sh~2-83 & & 4471, 4922, 5016, 6678 \\
Sh~2-100 & & 4026, 4471, 4922, 5016, 5048, 6678 \\
Sh~2-127 & & 4922, 5016, 6678 \\
Sh~2-128 & & 4471, 4922, 5016, 6678 \\
Sh~2-152 & & 4026, 4922, 5048, 5876, 7281 \\
Sh~2-209 & & 5876, 6678, 7281 \\
Sh~2-212 & & 4922, 5016, 5876 \\
Sh~2-288 & & 4471, 5016, 5048, 5876 \\
Sh~2-298 & & 4026, 4388, 4471, 4713, 4922, 5876 \\
Sh~2-308 & & 5876, 6678 \\
Sh~2-311 & & 3614, 3965, 4026, 4121, 4388, 4471, 4713, 4922, 5876, 6678, 7281 \\
\hline
\end{tabular}
\end{table*}

 \begin{table*}
 \centering
 \caption{Results of \texttt{Helio14} code: \ionic{He}{+}/\ionic{H}{+} ratios, minimal $\chi^2$ and  {\elect}(OII+OIII). }
\label{tab:list_helio14}
\begin{tabular}{lcccc}
\hline       
Object& Zone & \ionic{He}{+}/\ionic{H}{+} & $\chi^2$ & {\elect}(OII+OIII) (K)\\ 
\hline 
G2.4+1.4 & A1 & $10.800 \pm 0.132$ & 0.4824 & 8300 \\
& A2 & $10.913\pm	0.062$ &	0.2778&	9900 \\
& A3 & $10.823 \pm 0.025$ &0.0001&13200  \\
& A4 &  $10.911\pm	0.056$	&1.5670	&10800\\
& A5 &   $10.955	\pm0.104$&	2.1964&	9400\\
M8 & & $10.835\pm	0.006$&	3.1286&	8300\\
M16 & &  $10.892	\pm0.005$&	1.8101&	8200 \\
M17 & & $10.951	\pm0.010$&	1.7033&	8100 \\
M20 & & $10.849	\pm0.006$&	5.6603	&8200 \\
M42 & & $10.937\pm	0.005$&	10.4708	&8600 \\
NGC 2579  & & $10.932\pm	0.007$&	9.9796&	9100 \\
NGC 3576 & & $10.940	\pm0.005$&	6.4027&	8500 \\
NGC 3603 & & $10.997	\pm0.008$&	11.1381	&9200\\
NGC 6888 & A1 &  $10.938\pm	0.061$&	6.3886	&9000 \\
& A2 & $11.135\pm	0.010$&	3.6936	&8100\\
& A3 & $11.211\pm	0.024$&	1.1789	&8200 \\
& A4 & $11.248\pm	0.017$&	1.1147&	9800 \\
& A5 & $11.215\pm	0.031$&	0.4958&	9700 \\
& A6 & $11.222\pm	0.029$&	1.1959	&9300\\
RCW 52 & &$10.797\pm	0.135$&	0.1317	&7000 \\
RCW 58 & &  $11.218\pm	0.011$&	0.5340&	5700\\
NGC 7635 & A1 & $10.876	\pm0.010$&	0.2989	&7800 \\
& A2 & $11.049\pm	0.061$&	3.73   & 8000  \\
& A3 &  $10.979\pm	0.036$&	0.20  &  7900 \\
& A4 &  $10.794\pm	0.009$&	1.4294&	8900 \\
& A5 & $10.959\pm	0.016$&	2.2715&	7700  \\
& A6 & $10.926\pm	0.016$&	1.4503&	7500  \\
Sh~2-83 & &  $10.894\pm	0.026$&	0.0529&	10500\\
Sh~2-100 & & $10.969\pm	0.019$&	0.9411&	8300 \\
Sh~2-127 & & $10.804	\pm0.010$&	0.3323&	9800 \\
Sh~2-128 & & $10.927	\pm0.010$&	0.4683	&10200 \\
Sh~2-152 & & $10.794\pm	0.008$&	0.9671	&8100 \\
Sh~2-209 & & $10.855	\pm0.034$&	0.7623&	10700 \\
Sh~2-212 & & $11.003\pm	0.017$&	1.8543&	9200 \\
Sh~2-288 & & $10.792	\pm0.011$&	0.8347&	9400 \\
Sh~2-298 & & $10.967\pm	0.016$&	2.5547&	11700\\
Sh~2-308 & & $11.177\pm	0.056$&	0.0046&	16300\\
Sh~2-311 & &  $10.909\pm	0.006$&	0.9385&	9200\\
\hline
\end{tabular}
\end{table*}

\begin{table*}
\centering
\caption{ICF values.} 
\label{tab:icfs} 
\begin{tabular}{lcccc}
\hline
& \multicolumn{4}{c}{ICF(He) scheme} \\
Object & PTP77 & ZL03 & KS83 & B09   \\
\hline
G2.4+1.4  &  \multicolumn{4}{c}{-} \\
M8 &   $1.33 \pm 0.09$ & $1.15 \pm 0.02$  &  $1.23 \pm 0.07$  &  $1.82 \pm 0.36$ \\
M16 &  $1.59 \pm 0.21$ &  $1.38 \pm 0.08$ &  $1.24 \pm 0.07$  & $1.23 \pm 0.27$ \\
M17 &  $1.06 \pm 0.02$ &  $1.04 \pm 0.01$ &  $1.05 \pm 0.01$  &  $1.05 \pm 0.09$ \\
M20 &  $1.54 \pm 0.16$ &  $1.30 \pm 0.06$ &  $1.27 \pm 0.07$  &  $1.74 \pm 0.46$ \\
M42 &  $1.08 \pm 0.04$ &  $1.04 \pm 0.02$ &  $1.04 \pm 0.02$  &  $1.02 \pm 0.09$ \\
NGC 2579 &  $1.16 \pm 0.05$ &  $1.09 \pm 0.02$ &  $1.10 \pm 0.04$  &  $1.10 \pm 0.18$ \\
NGC 3576 &  $1.13 \pm 0.04$ &  $1.08 \pm 0.02$ &  $1.09 \pm 0.02$  &  $1.05 \pm 0.10$ \\
NGC 3603 &  $1.03 \pm 0.01$ & $1.02 \pm 0.01$ &  $1.02 \pm 0.01$  &  $1.03 \pm 0.13$ \\
NGC 6888 &   \multicolumn{4}{c}{-} \\
NGC 7635 A1 &  $1.37 \pm 0.17$ &  $1.23 \pm 0.09$ &  $1.22 \pm 0.08$  &  $1.32 \pm 0.46$ \\
NGC 7635 A2 &  $1.09 \pm 0.08$ &  $1.05 \pm 0.04$ &  $1.09 \pm 0.07$  &  $1.28 \pm 0.96$ \\
NGC 7635 A3 &  $1.10 \pm 0.10$ &  $1.05 \pm 0.05$ &  $1.10 \pm 0.08$  &  $1.39 \pm 1.31$ \\
NGC 7635 A4 &  $1.30 \pm 0.10$ &  $1.15 \pm 0.06$ &  $1.24 \pm 0.05$  &  $1.98 \pm 0.82$ \\
NGC 7635 A5 &  $1.20 \pm 0.23$ &  $1.12 \pm 0.13$ &  $1.15 \pm 0.14$  &  $1.21 \pm 1.05$ \\
NGC 7635 A6 &  $1.21 \pm 0.28$ &  $1.12 \pm 0.15$ &  $1.16 \pm 0.18$  &  $1.25 \pm 1.32$ \\
RCW 52 &  $1.39 \pm 0.49$ &  $1.22 \pm 0.17$ &  $1.27 \pm 0.36$  &  $2.12 \pm 2.47$ \\
RCW 58 &  $1.15 \pm 0.10$ &  $1.06 \pm 0.03$ &  $1.17 \pm 0.12$  &  $1.94 \pm 1.33$ \\
Sh~2-83 &  $1.04 \pm 0.02$ &  $1.05 \pm 0.02$ &  $1.02 \pm 0.01$  &  $1.25 \pm 0.45$ \\
Sh~2-100 &  $1.06 \pm 0.02$ &  $1.04 \pm 0.01$&  $1.05 \pm 0.01$  &  $1.04 \pm 0.13$ \\
Sh~2-127 &  $1.39 \pm 0.12$ &  $1.16 \pm 0.04$ &  $1.28 \pm 0.08$  & $1.47 \pm 0.63$ \\
Sh~2-128 &  $1.19 \pm 0.07$ &  $1.12 \pm 0.03$ &  $1.12 \pm 0.03$ &  $1.07 \pm 0.17$ \\
Sh~2-152 &  $1.31 \pm 0.07$ &  $1.09 \pm 0.02$ &  $1.27 \pm 0.05$  &  $3.35 \pm 0.64$ \\
Sh~2-209 &  $1.22 \pm 0.30$ &  $1.17 \pm 0.16$ &  $1.10 \pm 0.14$  &  $1.01 \pm 0.08$ \\
Sh~2-212 &  $1.28 \pm 0.40$ &  $1.16 \pm 0.15$ &  $1.20 \pm 0.30$  &  $1.50 \pm 1.60$ \\
Sh~2-288 &  $1.37 \pm 0.25$ &  $1.18 \pm 0.11$ &  $1.22 \pm 0.12$  &  $1.60 \pm 1.03$ \\
Sh~2-298 &  $1.91 \pm 0.53$ &  $1.95 \pm 0.20$ &  $1.15 \pm 0.06$  &  $1.46 \pm 0.40$ \\
Sh~2-308 &  $1.05 \pm 0.08$ &  $1.06 \pm 0.06$ &  $1.03 \pm 0.04$  &  $1.16 \pm 1.09$ \\
Sh~2-311 &  $1.47 \pm 0.14$ &  $1.33 \pm 0.05$ &  $1.23 \pm 0.07$  &  $1.20 \pm 0.18$ \\
\hline
\end{tabular}
\end{table*}

\bsp	
\label{lastpage}
\end{document}